\documentclass[11pt]{article}
\usepackage{graphicx}
\usepackage[margin=1.25in]{geometry}
\usepackage[usenames,dvipsnames]{color}
\usepackage{url}
\usepackage[colorlinks = true,
            linkcolor = blue,
            urlcolor  = blue,
            citecolor = blue,
            anchorcolor = blue]{hyperref}
\usepackage{hyperref,color}

\newcommand\pubnumber{FERMILAB-FN-1143-PPD-SCD }
\newcommand\pubdate{\today}

\def\Title#1{\begin{center} {\LARGE #1 } \end{center}}
\def\Author#1{\begin{center}{ \sc #1} \end{center}}
\def\Address#1{\begin{center}{ \it #1} \end{center}}

\newcommand\pubblock{\rightline{\begin{tabular}{l} \pubnumber\\
         \pubdate \end{tabular}}}
\newenvironment{Abstract}{\begin{quotation} \begin{center}
                       ABSTRACT
     \end{center}\bigskip  }{\end{quotation}}

\def\beq{\begin{equation}}
\def\eeq#1{\label{#1}\end{equation}}
\def\eeqn{\end{equation}}

\newenvironment{Eqnarray}%
   {\arraycolsep 0.14em\begin{eqnarray}}{\end{eqnarray}}
\def\beqa{\begin{Eqnarray}}
\def\eeqa#1{\label{#1}\end{Eqnarray}}
\def\eeqan{\end{Eqnarray}}

\let\bar=\overbar

\def\lsim{\mathrel{\raise.3ex\hbox{$<$\kern-.75em\lower1ex\hbox{$\sim$}}}}
\def\gsim{\mathrel{\raise.3ex\hbox{$>$\kern-.75em\lower1ex\hbox{$\sim$}}}}

\def\del{\partial}
\def\Dslash{\not{\hbox{\kern-4pt $D$}}}
\def\dslash{\not{\hbox{\kern-2pt $\del$}}}
\def\pslash{\not{\hbox{\kern-2pt $p$}}}
\def\ETmiss{\not{\hbox{\kern-4pt $E$}}_T}

\def\Dlr{\mathrel{\raise1.5ex\hbox{$\leftrightarrow$\kern-1em\lower1.5ex\hbox{$D$}}}}

\def\MSB{{\bar{M \kern -2pt S}}}
\def\msb{{\bar{\scriptsize M \kern -1pt S}}}

\def\drb{{\bar{\scriptsize D \kern -1pt R}}}

\newcommand\snowmass{\begin{center}\rule[-0.2in]{\hsize}{0.01in}\\\rule{\hsize}{0.01in}\\
\vskip 0.1in Submitted to the  Proceedings of the US Community Study\\ 
on the Future of Particle Physics (Snowmass 2021)\\ 
\rule{\hsize}{0.01in}\\\rule[+0.2in]{\hsize}{0.01in} \end{center}}

\begin{document}

\pubblock

\bigskip

\Title{Sensitivity to Dijet Resonances at Proton-Proton Colliders}

\bigskip 

\Author{Robert M. Harris}
\Address{Fermilab, P.O. Box 500, Batavia, IL, 60510, U.S.A.}

\medskip

\Author{Emine Gurpinar Guler, Yalcin Guler}
\Address{Konya Technical University, Road 42250, Selcuklu/Konya, Turkey}

\bigskip 

 \begin{Abstract}
\noindent 

A significant benchmark for discovery at a proton-proton collider is the sensitivity to a dijet resonance, X, the intermediate state of the s-channel process $pp \rightarrow X \rightarrow 2\mbox{ jets}$. To probe the highest resonance masses, hadron collider experiments have used the classic technique of searching for bumps in the mass spectrum of two individually resolved jets.  In this Snowmass 2021 study, we explore the search sensitivity to multiple benchmark models of dijet resonances at current and future proton-proton colliders. We present the expected masses for $5\sigma$ discovery or 95\% confidence level exclusion of diquarks, colorons, excited quarks, $W^{\prime}s$, $Z^{\prime}s$ and Randall-Sundrum gravitons, resulting from accumulation of integrated luminosities between 10 and $10^5$ fb$^{-1}$, at proton-proton colliders operating at energies $\sqrt{s}=$ 13, 14, 27, 75, 100, 150, 300 and 500 TeV.

\end{Abstract}

\snowmass

\clearpage

\tableofcontents

\clearpage

\def\thefootnote{\fnsymbol{footnote}}
\setcounter{footnote}{0}

\section{Introduction}

\begin{figure}[ht]
\begin{center}
\includegraphics[width=0.40\hsize]{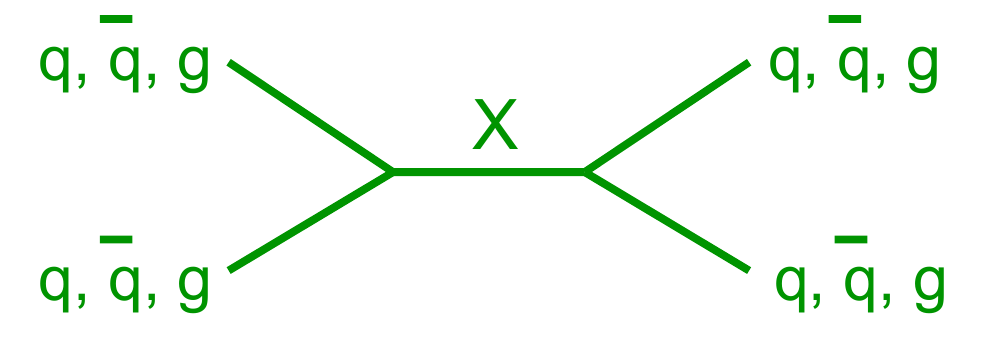}
\end{center}
\caption{s-channel production, via parton-parton annihilation, of a resonance (X) that subsequently decays to pairs of partons ($q$, $\bar{q}$ or $g$), giving a dijet in the final state.}
\label{figDiagram}
\end{figure}

Dijet resonances are an essential benchmark of discovery capability of proton-proton colliders.  The process, shown in Fig.~\ref{figDiagram}, is sensitive to a variety of new physics at the highest mass scales.  Predicted by many models of new physics that have been proposed to address fundamental questions, dijet resonances have been searched for by every hadron collider~\cite{review}. Proton-proton colliders are natural dijet resonance factories. Dijet resonances, X, produced by annihilation of partons in the colliding protons, must decay to two partons giving dijets.

This whitepaper estimates the sensitivity of pp colliders to this essential process. In the remainder of this section we discuss the benchmark models and pp collider scenarios we will consider. In section~\ref{secMethodology} we give the details of the methodology we use to evaluate the sensitivity of the pp collider scenarios. There we also present estimates of the QCD background, checked against the published results from the LHC, and present similar checks for the estimated sensitivities. In section~\ref{secSensitivity} we summarize the results of the study. We begin with examples of the discovery cross sections compared to the cross sections of the models. We then discuss examples of the discovery sensitivity at 5$\sigma$ significance and exclusion sensitivity at 95\% confidence level, for a few different ways to organize the models and pp collider scenarios.  In section~\ref{secConclusions} we conclude with some final observations.  In the appendix, section~\ref{secAppendix}, we tabulate the sensitivity results for all models and pp collider scenarios.

\subsection{Dijet Resonance Models}

We consider multiple benchmark models, exploring all of the parton-parton initial states available at pp colliders, and spanning a range of cross sections from various strengths of interaction and parton density. All models are narrow resonances, those for which the natural half-width, $\Gamma/2$, is smaller than the effective experimental resolution. We perform lowest order calculations, using the same code as used by ref~\cite{refCMS}, of the product of the cross section, branching ratio, and acceptance in the narrow-width approximation, using CTEQ6L1 parton distributions and renormalization scale equal to the resonance mass. We classify the models into two groups according to their production cross section: strongly and weakly produced.

\begin{figure}[ht]
\begin{center}
\includegraphics[width=0.9\hsize]{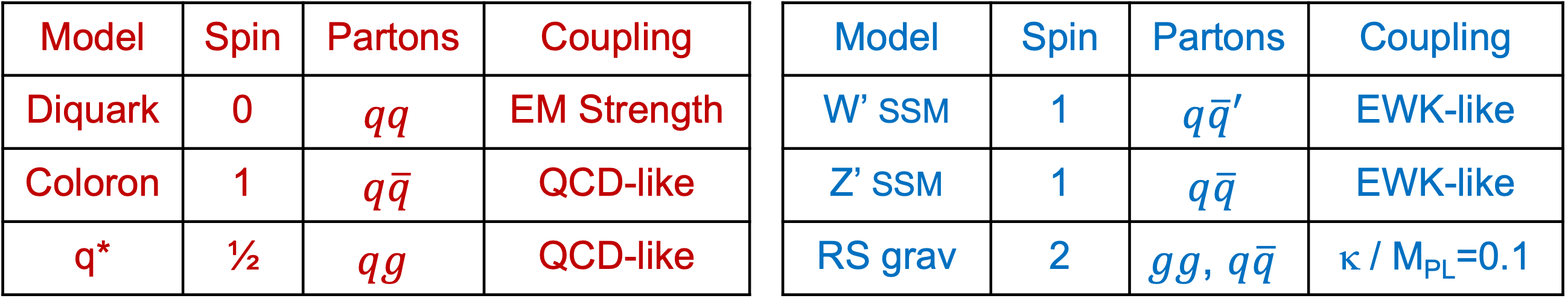}
\end{center}
\caption{Benchmark models of narrow resonances. Strongly produced models (left), include scalar diquarks, colorons and excited quarks. Weakly produced models (right), include heavy bosons W' and Z', and the Randall-Sundrum graviton. The tables list the resonance's spin, the partons within the resonance production and decay process, and the coupling strength.}
\label{figModels}
\end{figure}

\subsubsection{Strongly Produced Resonances}

Figure~\ref{figModels} summarizes the models. The strongly produced models are the ones that have larger cross section, typically of order the QCD background cross section.  The largest cross section are for the scalar diquarks~\cite{refDiquark}, despite couplings to $uu$ and $dd$ assumed to be electromagnetic strength. This is because diquarks are produced from large parton densities, the parton distribution functions (PDF), found for the valence $u$ and $d$ quarks inside each of the colliding protons. Colorons are massive gluons from an extra color interaction, where we assume a minimal mixing betweeen the two color interactions ($\cot\theta=1$). For this choice of coupling the resonance is narrow, and the production cross sections are identical for two canonical models of colorons: axigluons~\cite{refAxigluon} and flavor-universal colorons~\cite{refColoron}. Also produced from the color-like interaction, specifically a chromo-magnetic transition, are excited quarks ($q^*$) predicted by models of quark compositeness~\cite{refQstar}. We consider mass-degenerate first generation excited quarks ($u^*$, $d^*$), standard model couplings ($f=f_s=f^{\prime}=1$), and set the energy scale $\Lambda$ of quark compositeness equal to the mass of the excited quark.  While this is a conventional model that has been used by every search at hadron colliders, we note that this model implies an even larger effect from the contact interaction of the colliding quarks at this compositeness scale~\cite{refContact}. Such an effect should be experimentally observed earlier than a resonance, from measurements of the shape of the jet $p_T$, dijet mass, or dijet angular distributions.

\subsubsection{Weakly Produced Resonances}

The weakly produced models in Fig.~\ref{figModels} are the ones having smaller cross section, typically of order the electroweak processes.  The new gauge bosons $W^{\prime}$ and $Z{^\prime}$~\cite{refGauge}, are the canonical heavy cousins to the electroweak bosons $W$ and $Z$, respectively. They have the same couplings to quarks and leptons as in the standard model, and hence the model is frequently called the sequential standard model (SSM). When calculating the branching fraction to dijets we only include decays to the five lightest quarks, so the decays to top quarks reduces our branching fraction to jets, and we neglect possible decays to $W$ or $Z$ bosons. A next-to-leading order correction factor of $K=1+ 8\pi\alpha_s/9$ is applied to the leading order predictions for the $W^{\prime}$ model and $K=1+(4\alpha_/6\pi)(1+4\pi^2/3)$ for the $Z^{\prime}$ model. 
Randall--Sundrum (RS) gravitons are predicted in a model of warped extra dimensions~\cite{refRSG}. The value of the dimensionless coupling parameter $k/\overline{M}_{Pl}$ is chosen to be 0.1, as is usually done for dijet resonance searches. RS graviton cross sections are typically the lowest of all the models considered, so we classify them as weakly produced, although we note that the production mechanism has nothing to do with electroweak interactions.

\subsection{pp Colliders}
\label{secColliders}

We perform a comprehensive study of all scenarios for current and future pp colliders, considering eight collision energies ($\sqrt{s}$): 
\begin{enumerate}
\item 13 TeV: Large Hadron Collider (LHC) during Run 2 occurring from 2016 to 2018.
\item 14 TeV: LHC in the High Luminosity (HL-LHC) era.
\item 27 TeV: High Energy LHC (HE-LHC) and Fermilab site filler (FNAL-SF) proton-proton options.
\item 75 TeV: Super Proton-Proton Collider (SPPC) and a lower energy option for the Future Circular Collider-hadron hadron (FCC-hh).
\item 100 TeV: Median and usually assumed energy for FCC-hh.
\item 150 TeV: High energy option for FCC-hh or a similar machine.
\item 300 TeV: A Very Large Hadron Collider (VLHC) proposed at a snowmass energy frontier workshop.
\item 500 TeV: Collider in the Sea, also discussed at snowmass energy frontier workshop.
\end{enumerate}

We consider ten integrated luminosities ($\int L\/dt$). Five general decades, $10^1$-$10^5$ fb$^{-1}$, and five baseline values which were actually used for LHC runs or anticipated for future runs and colliders : 140 fb$^{-1}$ (LHC Run 2), 200 fb$^{-1}$ (LHC Run 3), 3 ab$^{-1}$ (HL-LHC), 2.5 and 30 ab$^{-1}$ (FCC-hh).

We determine the mass sensitivity for discovery and exclusion of each dijet resonance model at all values of collision energy and integrated luminosity.

\section{Methodology}
\label{secMethodology}

In order to obtain sensitivities for six models, considering eight collision energies and ten integrated luminosities, we require a simple and fast methodology. We adopt and update the methodology of a prior study one of the authors had conducted for Snowmass 1996~\cite{Snowmass96}, using lowest order parton level calculations of the QCD background and the dijet resonance signal. QCD calculations are performed using CTEQ6L1 parton distributions with a renormalization scale of $\mu=p_T/2$, and dijet resonance calculations are performed with the same PDF and a renormalization scale equal to the resonance mass.

Selection cuts are similar to publications from LHC~\cite{refCMS,refATLASdijet}. The dijet mass, $m$, is calculated for the two final state partons with pseudo-rapidity $|\eta|<2.5$. To suppress the large background from QCD t-channel scattering of the partons, we require the pseudorapidity separation of the the two partons to satisfy $|\Delta\eta|<1.1$, which is equivalent to selecting events with a center-of-momentum frame scattering angle $|\cos\theta^*|<0.5$. 

The expected number of signal and background events, inside a search window in $m$, is calculated and used to estimate sensitivities.  As in the prior Snowmass 1996 study, the window is centered on the pole mass of the resonance, $M$, and has a width equal to 16.4\%: $0.836 M < m < 1.164 M$.  That width, originally constructed to provide 90\% acceptance for a Gaussian with width 10\% of the pole mass, turns out to accept a reasonable amount of signal for the dijet resonance shapes we expect, discussed in section~\ref{secAcc}.

From the estimated number of background events in the mass window, $N_{QCD}$, we can estimate the expected exclusion or discovery cross section.  

The 95\% confidence level excluded number of events is defined as 
\begin{equation}
N_{excl} = (3 + \sqrt{3^2 + 4(1.64^2)N_{QCD}})/2
\label{eq95CL}
\end{equation}
When $N_{QCD}$ is large, this equation predicts the known number of signal events that can be excluded at 95\% CL using Gaussian statistics, $N_{excl}=1.64\sqrt{N_{QCD}}$. When $N_{QCD}=0$, it also gives the known number of events for 95\% exclusion from Poisson statistics, $N_{excl}=3$ events. We have checked that Eq.~\ref{eq95CL} also (remarkably) gives the exact Poisson statistical answer within 0.5\% accuracy for the intermediate region, at least for $0 \leq N_{QCD} \leq 10$.

The 5 $\sigma$ discovery number of events is defined as 
\begin{equation}
N_{5\sigma} = (25 + \sqrt{25^2 + 4(5^2)N_{QCD}})/2 
\label{eq5sigma}
\end{equation}
When $N_{QCD}$ is very large, as is usually the case, this equation predicts the known number of events for discovery from Gaussians statistics, $N_{5\sigma}=5\sqrt{N_{QCD}}$. When $N_{QCD}=0$, it conservatively assigns the number of events required for discovery at $N_{5\sigma}=25$, which is the events required to measure a cross section at $5\sigma$. Equation~\ref{eq5sigma} smoothly connects the regions of high and low statistics in the same way as Eq.~\ref{eq95CL} did.

\begin{figure}[hbt]
\begin{center}
\includegraphics[width=0.49\hsize]{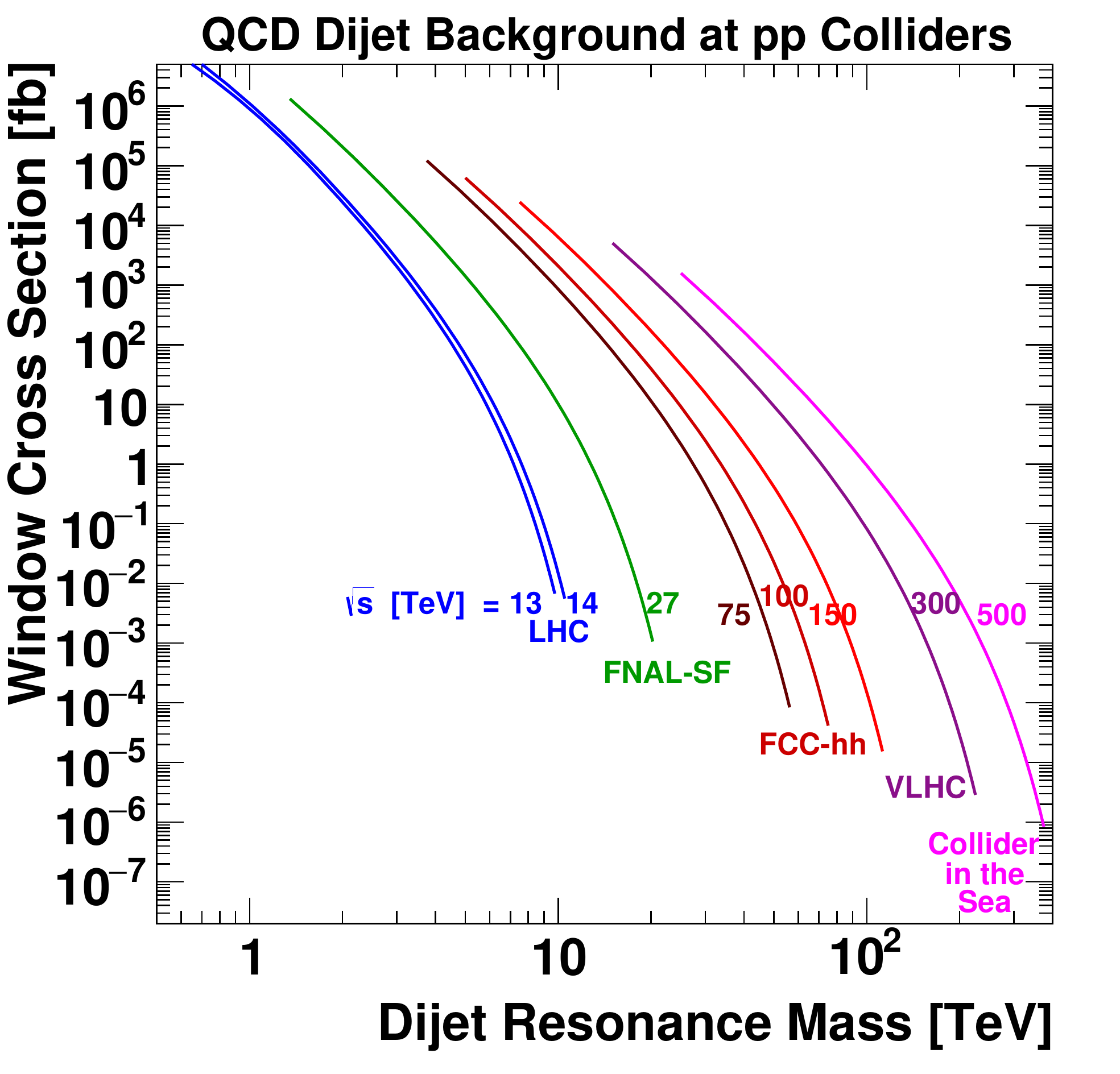}
\includegraphics[width=0.49\hsize]{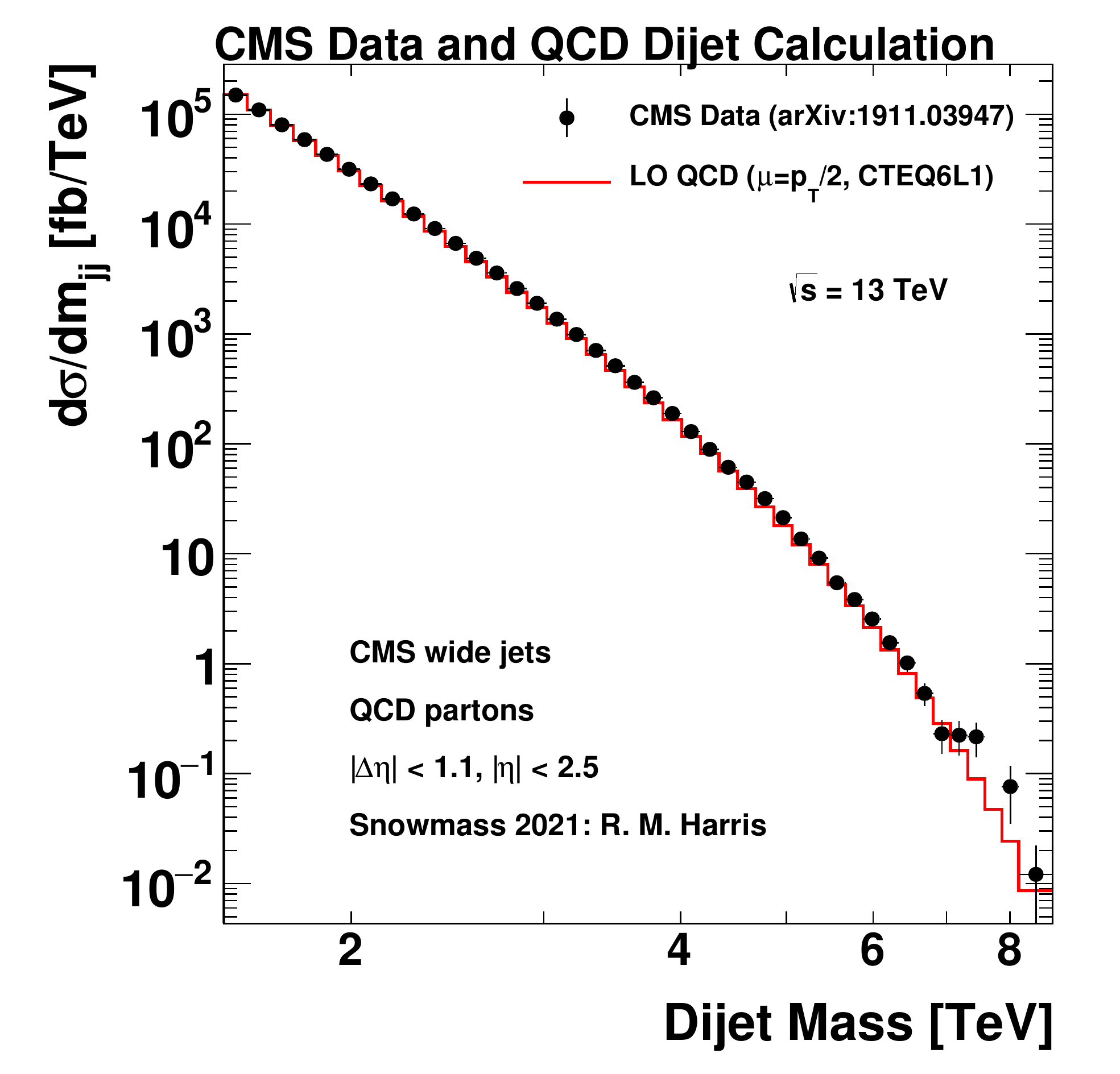}
\end{center}
\caption{Left) The QCD background as a function of dijet resonance mass, from a lowest order parton level calculation of the cross section within a 16.4\% mass window centered on the resonance pole, is shown for all collider energies from 13 to 500 TeV. Right) For $\sqrt{s}=13$ TeV, the lowest order parton level QCD calculation of the differential cross section as a function of dijet mass (histogram), is compared to published data from Ref.~\cite{refCMS} (points), as a check of the calculation on the left.}
\label{figQCD}
\end{figure}

\subsection{QCD Background}

In Fig.~\ref{figQCD} we show the QCD background cross section to a dijet resonance for all scenarios of pp colliders considered, from the LHC to the Collider in the Sea.  The QCD background is calculated for resonance masses between 5\% and 75\% of the collision energy, which covers the typical range for a high mass search at a hadron collider. We note that the QCD background at a mass proportional to the collision energy decreases smoothly with increasing collision energy. Figure~\ref{figQCD} also shows that the QCD calculation at $\sqrt{s}=13$ TeV agrees with the measured CMS dijet data~\cite{refCMS} to within about 10\%. This is likely because CMS uses the wide jet algorithm,  that corresponds well to the partons within a $2\rightarrow 2$ process.  We have also explicitly chosen the renormalization scale $\mu=p_T/2$ for all colliders, because it gives a larger normalization of the QCD cross section, which is closer to the observed CMS data than the choice $\mu=p_T$ used in the previous study~\cite{Snowmass96}. Thus, while our calculations are only at lowest order, our ability to predict existing pp collider data increases our confidence in the QCD background estimates at future pp colliders. 

\subsection{Signal Acceptance}
\label{secAcc}

\begin{figure}[hbt]
\begin{center}
\includegraphics[width=0.48\hsize]{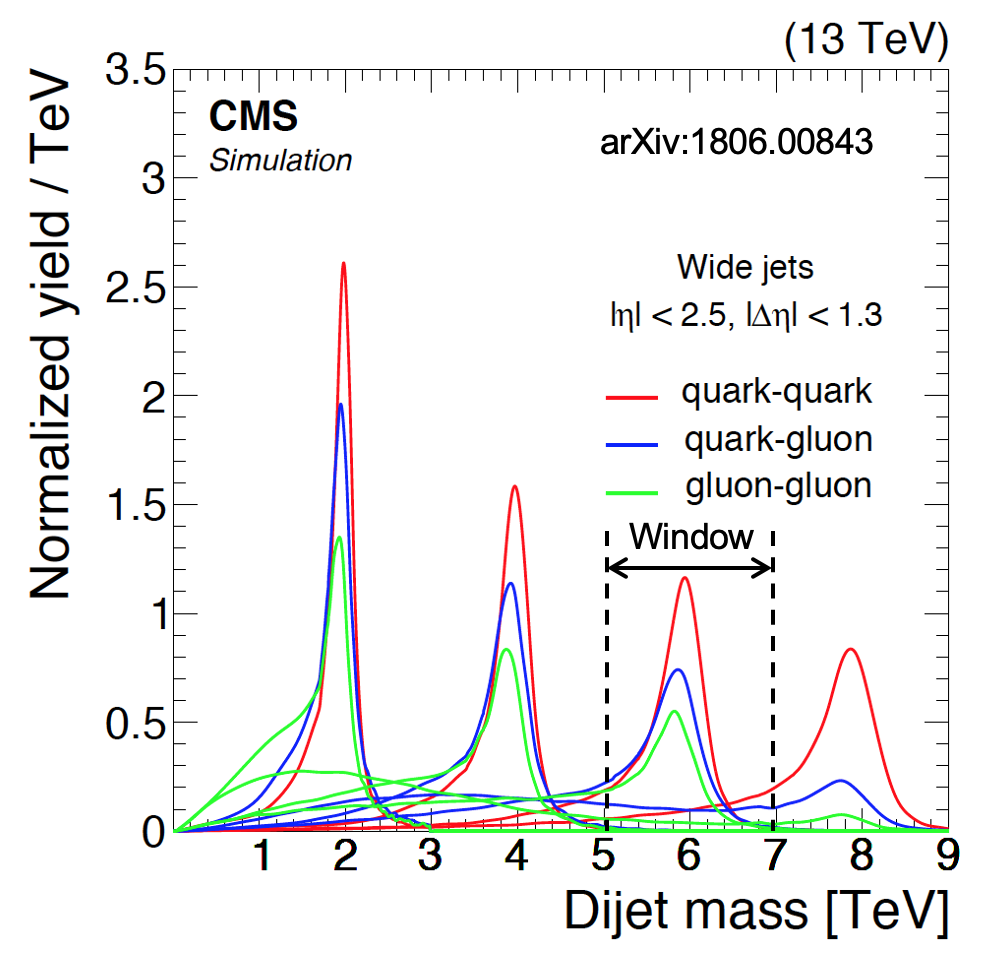}
\includegraphics[width=0.5\hsize]{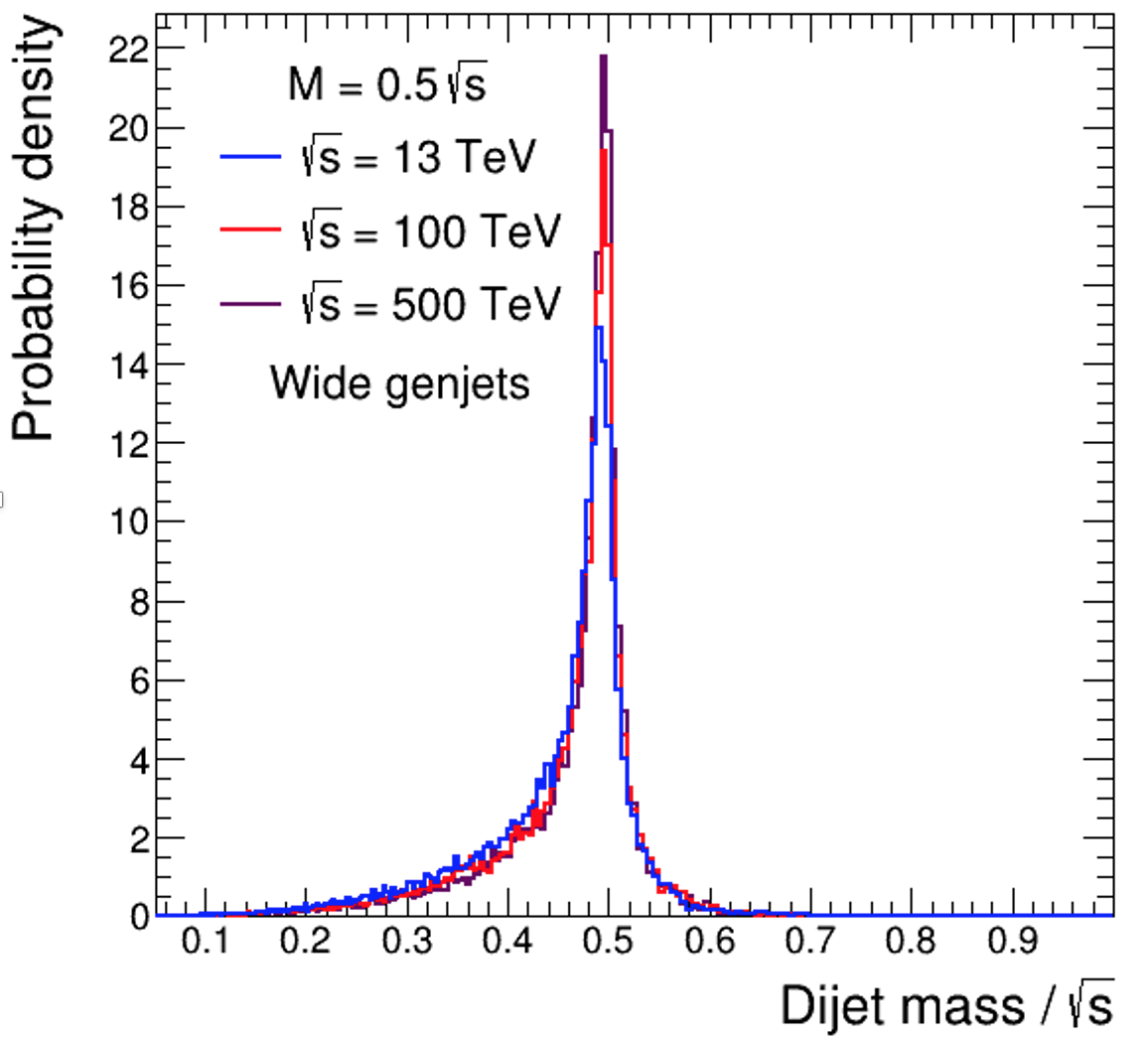}
\end{center}
\caption{Left) Experimental simulations of fully reconstructed dijet resonance signal shapes at $\sqrt{s}=13$ TeV from CMS~\cite{CMS:2018mgb} with our search window superimposed. Right) Simulations of dijet resonance signal shapes at genjet level, for excited quarks with a mass equal to $0.5\sqrt{s}$, are compared for three collision energies: $\sqrt{s}=13$, $100$ and $500$ TeV. For both plots the wide jet algorithm is used. }
\label{figAcceptance}
\end{figure}

The signal acceptance of our mass window for the fully reconstructed shape of a dijet resonance is illustrated in Fig.~\ref{figAcceptance}, which shows signal simulations using PYTHIA. The fully simulated and reconstructed resonance shape from ref.~\cite{CMS:2018mgb} at $\sqrt{s}=13$ TeV, has an approximately Gaussian core from experimental resolution, and a long tail to low values of dijet mass. The tail is primarily caused by final state radiation falling outside the reconstructed jet. To minimize this tail and optimize the signal significance ref.~\cite{CMS:2018mgb} used the wide jet algorithm to reconstruct dijets, where the two jets with the highest transverse momentum, reconstructed with the anti-$k_T$ algorithm and a distance parameter of 0.4 (AK4 jets), are used as seeds for the wide jets which contain selected AK4 jets within $\Delta R=\sqrt{(\Delta\eta)^2 + (\Delta\phi)^2}<1.1$. This effectively widens the cone size to 1.1, recovering the energy lost to jets radiated outside that distance, with transverse momentum $p_T > 30$ GeV inside the fiducial region $|\eta|<2.5$. Applying our window to the fully reconstructed resonance shapes at $\sqrt{s}=13$ TeV, as illustrated in Fig.~\ref{figAcceptance}, gives the acceptance for each of the basic types of resonances: quark-quark (qq), quark-gluon (qg) and gluon-gluon (gg).


The acceptance for any resonance type should be approximately the same at all $\sqrt{s}$. First, this is based on the observation that the generated shape of a dijet resonance, with pole mass equal to a fixed fraction of $\sqrt{s}$, is indeed approximately independent of $\sqrt{s}$. Figure~\ref{figAcceptance} demonstrates the approximate invariance of the shape of a dijet resonance for the full range of collision energies we consider. We are using particle level quantities from the generator, with wide genjets, where we reconstruct AK4 genjets and cluster them into wide jets if they pass the aforementioned $|\eta|$ and $\Delta R$ cuts, and a generalized transverse momentum cut scaled to the collider energy: $p_T>(\sqrt{s}/13\ \mbox{TeV})30$ GeV. Second, we reason that any detector will need to be designed to deliver an appropriate experimental resolution for jets, and that resolution would be one that delivers at least a similar intrinsic Gaussian core resolution as previous experiments at proton-proton colliders. We further note that the natural range of dijet mass exploration scales with $\sqrt{s}$ in the same was as do the genjet resonances shown in Fig.~\ref{figAcceptance}.  Thus, the window acceptances of the fully simulated and reconstructed shapes of ref.~\cite{CMS:2018mgb}, should be approximately applicable to any $\sqrt{s}$ value. Furthermore, the signal acceptances, shown in Fig.~\ref{figLimits13TeV}, depend more strongly on the type of resonance (qq, qg and gg), than on the resonance mass value for a given type, so we have used the acceptances tabulated in Fig.~\ref{figLimits13TeV} as a function of model for all values of $\sqrt{s}$, constant as a function of resonance mass. See section~\ref{secAppendix} for more discussion of the accuracy of this approximation.

\subsection{Check: Limits at LHC Run 2}

\begin{figure}[hbt]
\begin{center}
\includegraphics[width=0.44\hsize]{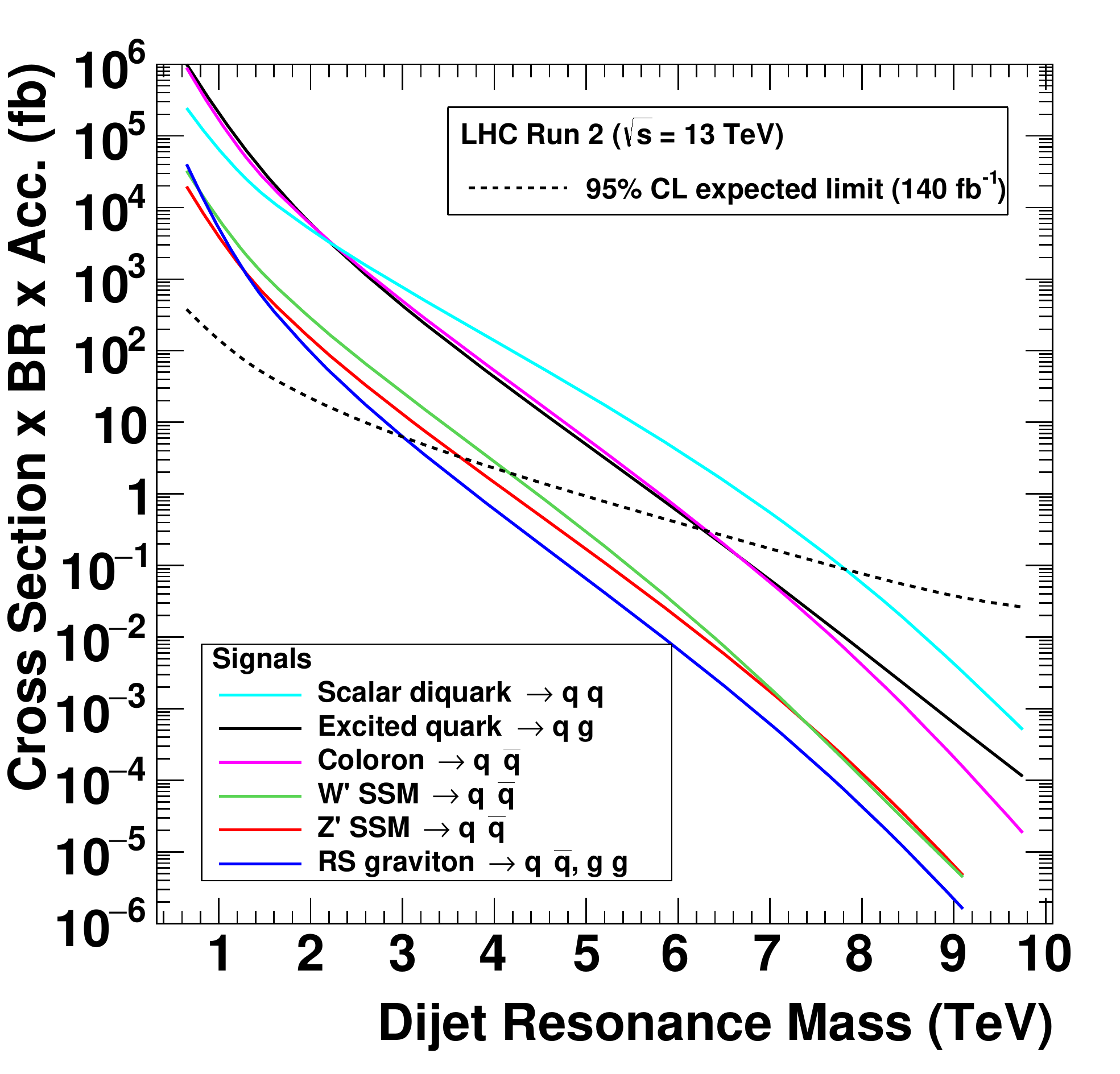}
\includegraphics[width=0.54\hsize]{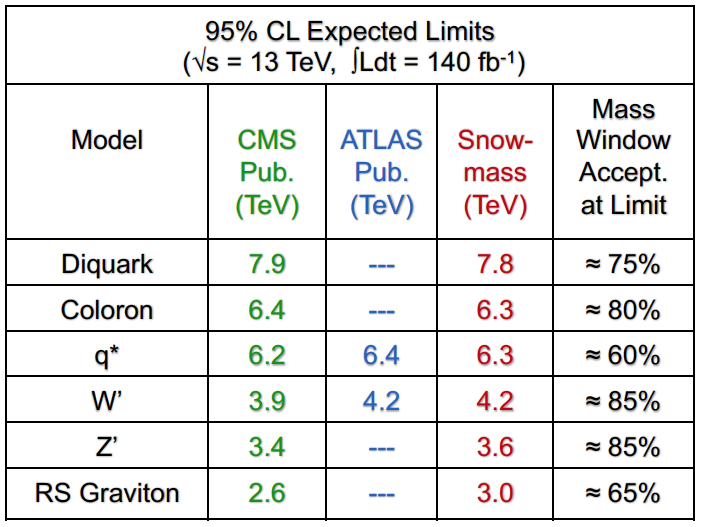}
\end{center}
\caption{Limits at the LHC during Run 2. Left) Our estimation using this analysis of the 95\% CL expected limit (dashed curve) on the product of cross section, branching fraction and acceptance, is compared with the signal prediction for six models of dijet resonances (solid curves) at $\sqrt{s}=13$ TeV for 140 pb$^{-1}$. Right) For the same models, the published 95\% CL expected mass limits from the LHC experiments (CMS~\cite{refCMS}, ATLAS~\cite{refATLASdijet}) are compared to the mass limits from this analysis (Snowmass). The acceptance of the mass window for the CMS resonance shapes in Fig.~\ref{figAcceptance} are also shown for the mass value at which the limit is found on a given model. }
\label{figLimits13TeV}
\end{figure}

The LHC experiments provide an example pp collider scenario, a testbed for our methodology, and an important check. In Fig.~\ref{figLimits13TeV} we show for $\sqrt{s}=13$ TeV the expected upper limits on the QCD background cross section within the mass window, compared to the signal cross section within that mass window. The signal cross section is the product of the full signal cross section, branching fraction to dijets, angular acceptance of the jet $|\eta|<2.5$ and $|\Delta\eta|<1.1$ cuts which is a part of our calculation, and the mass window acceptance tabulated in the figure.  The expected mass limit on each model is equal to the mass value at which the expected upper limit on the window cross section is equal to the signal cross section for the model. Fig.~\ref{figLimits13TeV} explicitly demonstrates that the expected mass limit on each model of dijet resonance is approximately equal to the published limits of the LHC experiments~\cite{refCMS, refATLASdijet}. We conclude that the methodology of this analysis, estimating dijet resonance sensitivity from a lowest order calculation of events observed within a mass window, provides sufficient accuracy.

\clearpage

\subsection{Determination of Sensitive Cross Section and Mass}
\label{secMetrics}
\begin{figure}[hbt]
\begin{center}
\includegraphics[width=0.49\hsize]{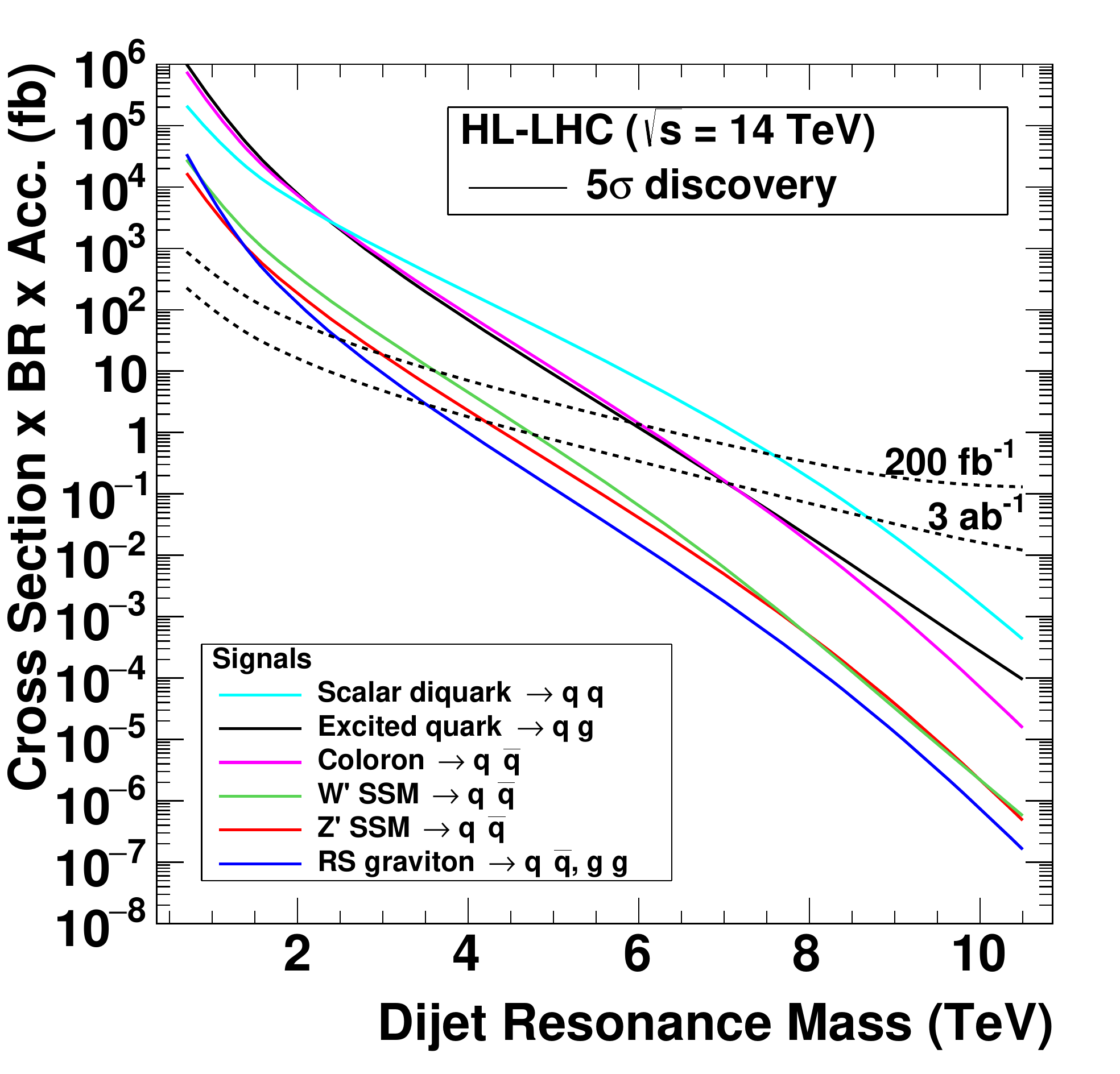}
\includegraphics[width=0.49\hsize]{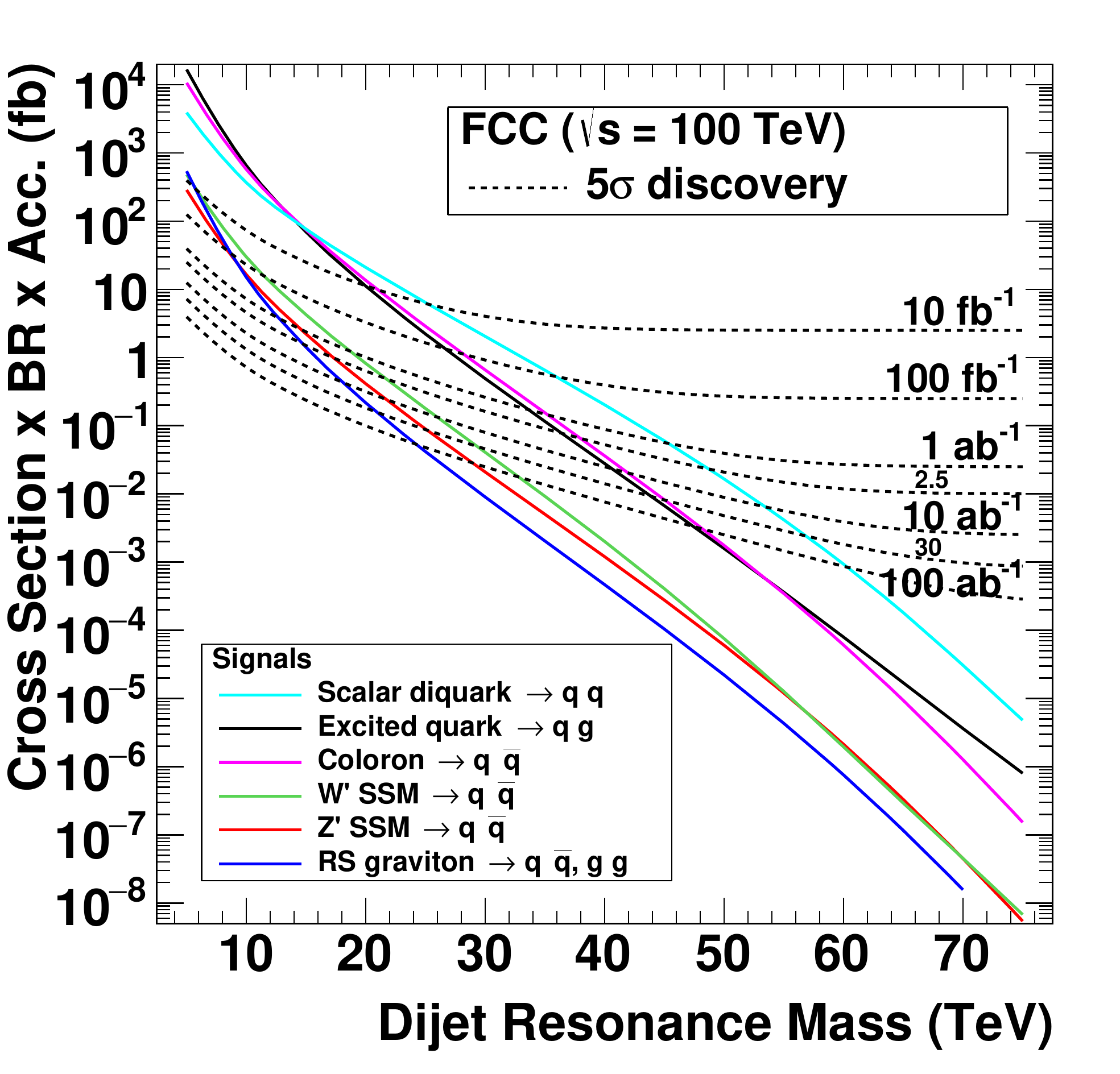}
\end{center}
\caption{ Discovery cross sections and models. The $5\sigma$ discovery value on the product of cross section, branching fraction and acceptance, for the integrated luminosities shown (dashed curves), is compared with the signal prediction for six models of dijet resonances (solid curves). Left) For HL-LHC at $\sqrt{s}=14$ TeV. Right) For FCC-hh at $\sqrt{s}=100$ TeV.}
\label{figDiscoveryXsec}
\end{figure}

The $5\sigma$ discovery cross section, the number of events required to make a $5\sigma$ discovery divided by the integrated luminosity, and the model cross sections times acceptance in the mass window, are calculated for all pp collider scenarios discussed in section~\ref{secColliders}.  The $5\sigma$ discovery masses on a given model are those mass values for which the discovery cross section and the model cross section are equal.  This is illustrated for HL-LHC and FCC-hh in  Fig.~\ref{figDiscoveryXsec}, where we compare the discovery cross sections with the corresponding signal cross sections for models of dijet resonances. The discovery masses are the masses where the curves cross. This process is repeated for all collider scenarios. Note that the discovery cross section is inversely proportional to the square root of the integrated luminosity when the background is large, and Gaussian statistics dominates, and is inversely proportional to the integrated luminosity when there is no background and discovery requires a fixed number of events.  

The process is similar for the determination of 95\% CL limits on the cross section and mass, which is also performed for every pp collider scenario in section~\ref{secColliders}. We have illustrated the process of finding these limits for the case of the LHC in Fig.~\ref{figLimits13TeV}. The scaling behavior of the cross section limits with integrated luminosity is the same as noted above for discovery cross sections.

\section{Mass Sensitivity}
\label{secSensitivity}

We use the $5\sigma$ discovery mass and the 95\% CL mass limits, defined in section~\ref{secMetrics} and listed in Tables~\ref{tabDiscoveryStrong}-\ref{tabLimitWeak} in section~\ref{secAppendix}, as metrics of sensitivity to dijet resonances at proton proton colliders. They are discussed for various combinations of our benchmark models and machine scenarios in this section.

\subsection{$q^*$ and $Z^{\prime}$}

\begin{figure}[hbt]
\begin{center}
\includegraphics[width=0.49\hsize]{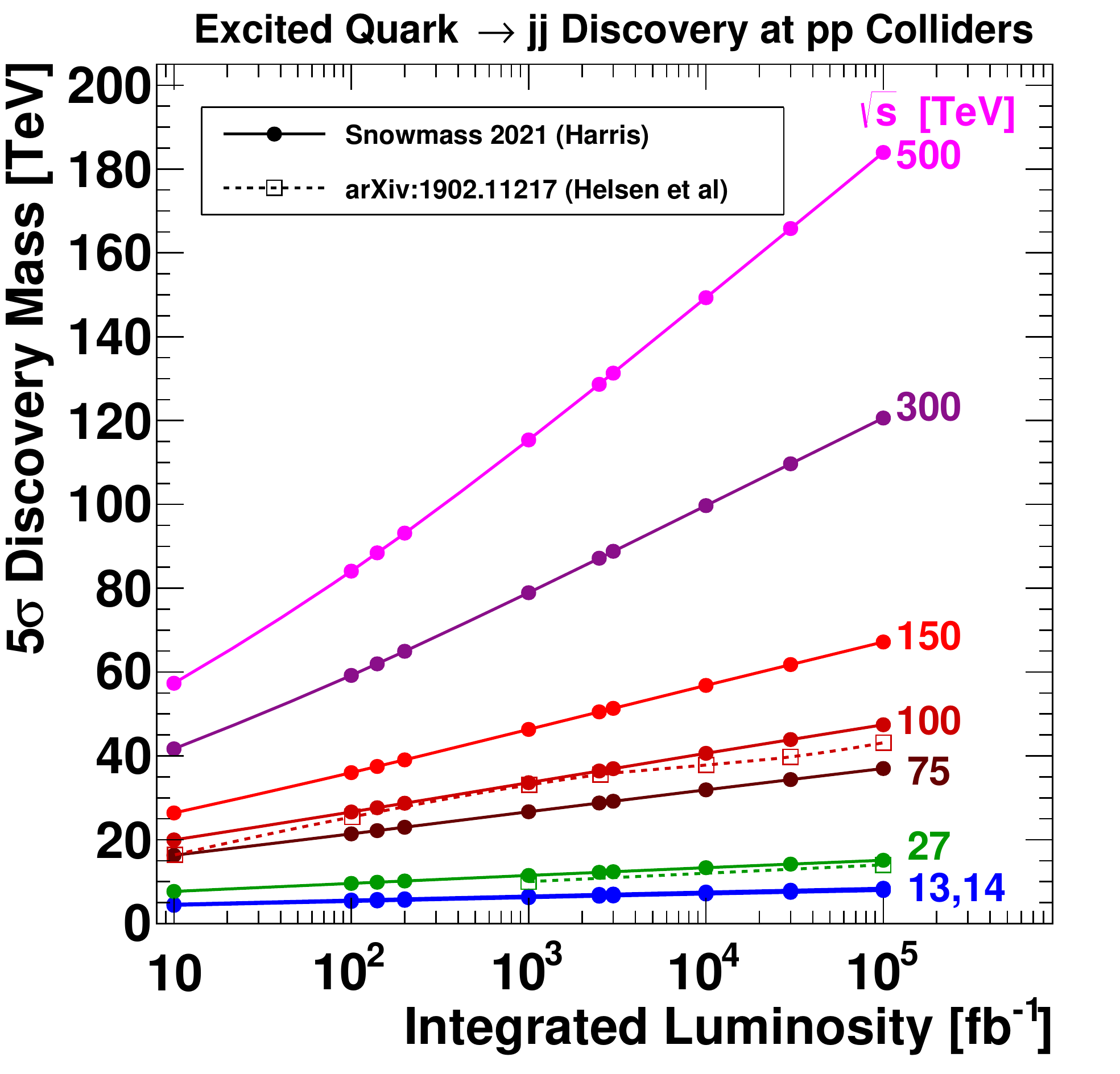}
\includegraphics[width=0.49\hsize]{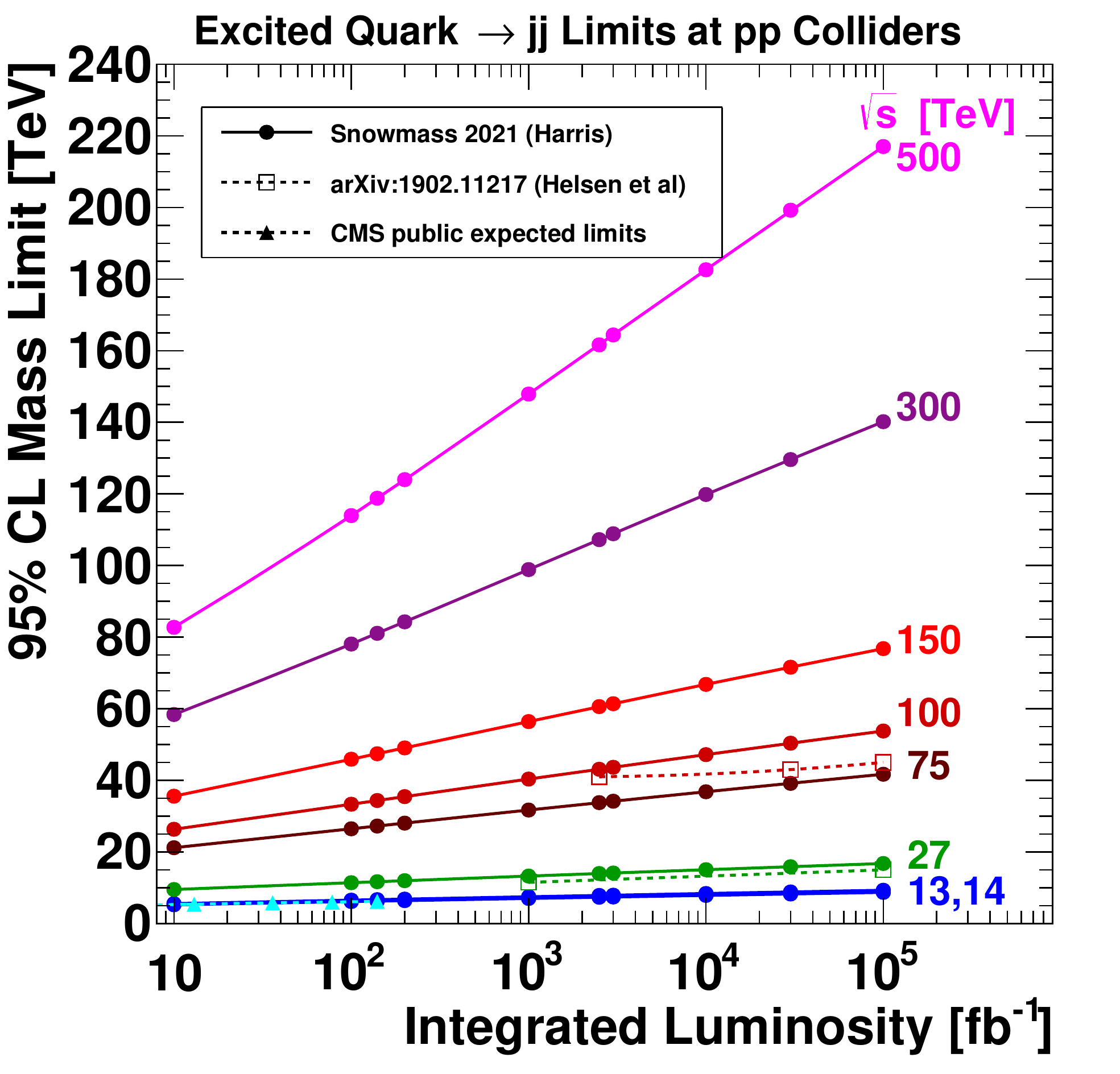}
\end{center}
\caption{ Sensitivity to an excited quark. Left) the $5\sigma$ discovery mass and Right) the 95\% CL mass limit, for eight values of collider $\sqrt{s}$ (colors), at ten values of integrated luminosity (solid circles). Also shown for comparison are both types of sensitivities for $\sqrt{s}=27$ and $100$ TeV from the analysis of Ref.~\cite{Helsens:2019bfw} (open boxes), and the public expected mass limits for $\sqrt{s}=13$ TeV from CMS~\cite{refCMS,CMS:2018mgb,CMS:2016gsl,CMS:2018wxx} (solid stars).}
\label{figQstar}
\end{figure}

\begin{figure}[hbt]
\begin{center}
\includegraphics[width=0.49\hsize]{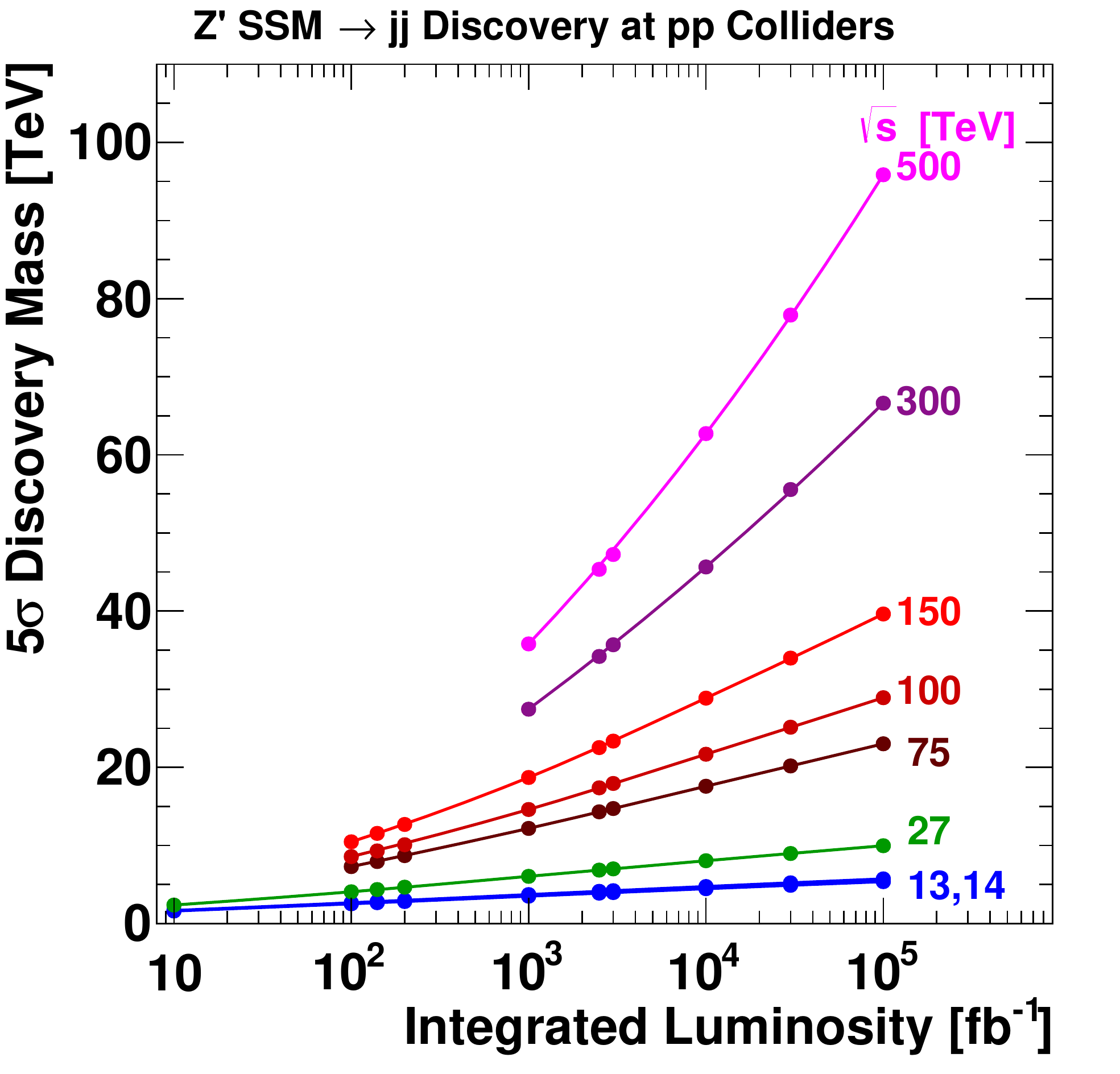}
\end{center}
\caption{ the $5\sigma$ discovery mass for an SSM $Z^{\prime}$, for eight values of collider $\sqrt{s}$ (colors), at multiple values of integrated luminosity (solid circles).}
\label{figZprime}
\end{figure}

The sensitivity to an excited quark at proton proton colliders is shown in Fig.~\ref{figQstar}. This is the most frequently used benchmark model for dijet resonances at hadron colliders, and was the first model searched for by both CMS and ATLAS when the LHC was turned on. Fig.~\ref{figQstar}  demonstrates that the sensitivity results from this analysis are close to those of previous studies, and in excellent agreement with limits from CMS.

We note two scaling behaviors of the sensitivity. First, the mass sensitivity is proportional to the collision energy. This is expected as the parton energies are proportional to the collision energies. Second, for any fixed value of $\sqrt{s}$, the sensitivity is proportional to the logarithm of the integrated luminosity. Again this is expected, as significant gains in sensitivity require orders of magnitude more luminosity, due to the exponential falloff of the parton distributions as a function of energy. 

The sensitivity of a Z$^{\prime}$ boson in the SSM at proton-proton colliders is shown in Fig.~\ref{figZprime}. This is the most frequently used benchmark of a heavy boson which is weakly produced. The results shown in Fig.~\ref{figZprime} are only reported for resonance mass values greater than $0.06\sqrt{s}$, the lowest mass for which our window lower edge is above our QCD background minimum dijet mass of $0.05\sqrt{s}$. Similar to excited quarks, the mass sensitivity of $Z^{\prime}$ bosons is approximately proportional to $\sqrt{s}$ and to the logarithm of integrated luminosity.

\clearpage

\subsection{Strongly and Weakly Produced Models}

\begin{figure}[hbt]
\begin{center}
\includegraphics[width=0.49\hsize]{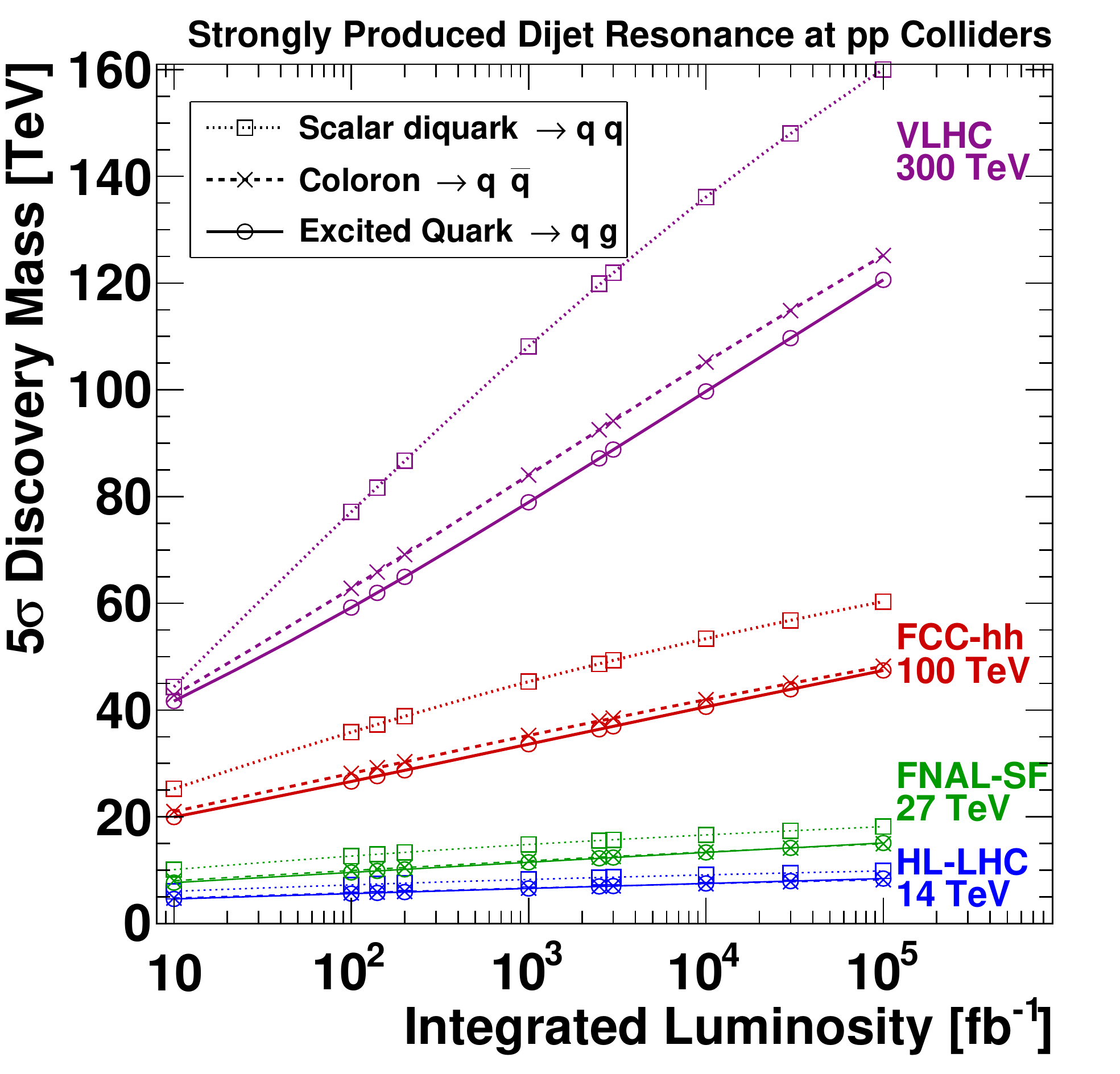}
\includegraphics[width=0.49\hsize]{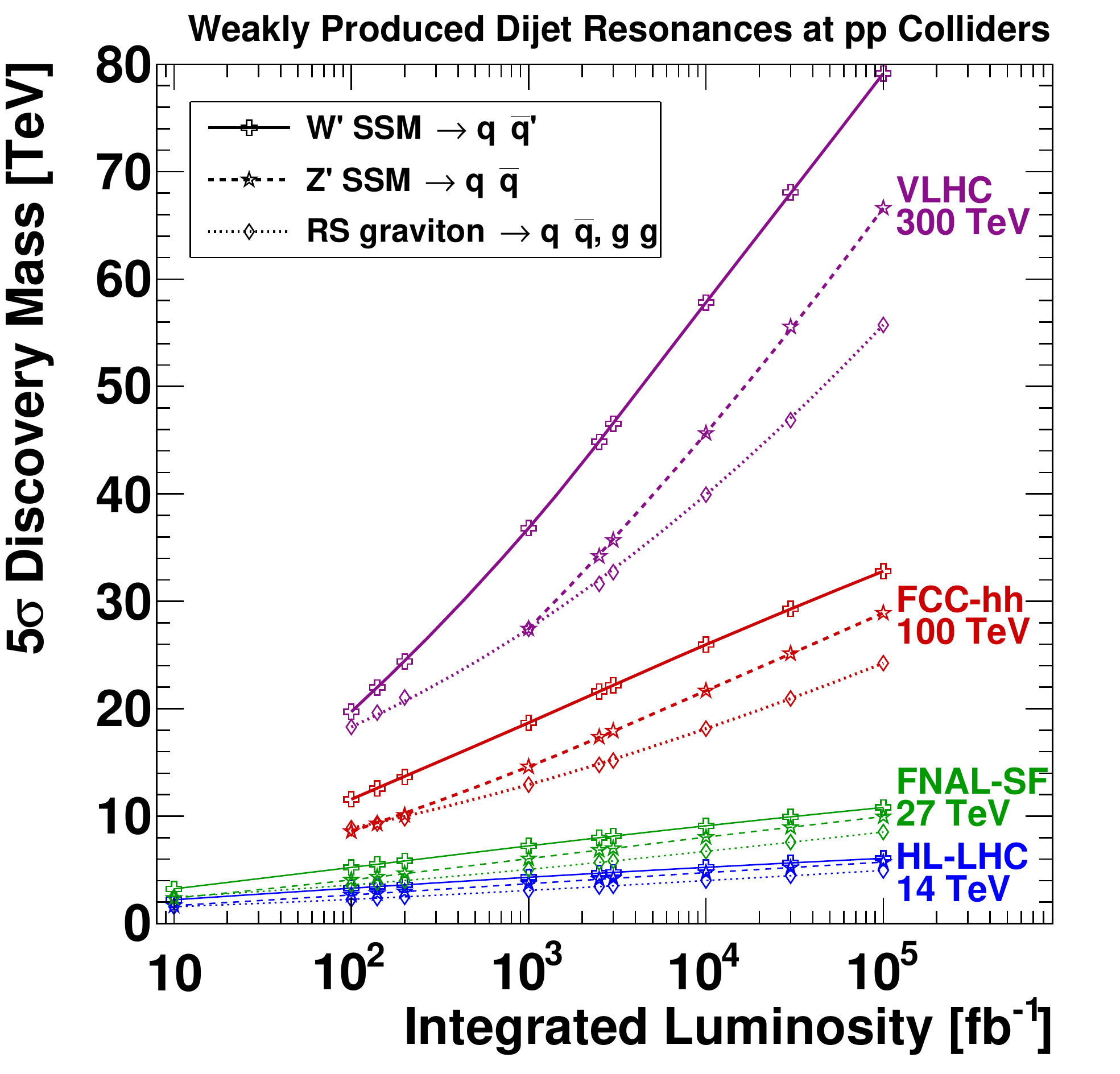}
\end{center}
\caption{ Sensitivity to strongly and weakly produced resonance models. The $5\sigma$ discovery mass for four values of collider $\sqrt{s}$ (colors) as a function of integrated luminosity for dijet resonances from (left) the large cross section models of diquarks (boxes), colorons (Xs), and excited quarks (circles) and from (right) the smaller cross section models of W$^{\prime}$ SSM bosons (crosses)  $Z^{\prime}$ SSM bosons (stars) and Randall-Sundrum gravitons (diamonds). }
\label{figStrongWeak}
\end{figure}

The sensitivity to a dijet resonance is mainly determined by it's cross section, which is why we present the sensitivity organized in two sets of models in Fig.~\ref{figStrongWeak}, strongly produced models with a cross section that is roughly two orders of magnitude greater than weakly produced models.  The maximum vertical axis value in the two plots in Fig.~\ref{figStrongWeak} shows that the discovery mass for the strongly produced resonances is roughly twice that for the weakly produced resonances. Among the strongly produced models the sensitivity is largest for diquarks, because the valence quark distribution of the proton becomes the largest PDF when the momentum of the quark approaches and exceeds one third of the momentum of the proton. The coloron requires an anti-quark to produce it, and the excited quark requires a gluon, and the PDFs for these partons are significantly smaller than for quarks at high momentum. Among the weakly produced models the sensitivity is largest for $W^{\prime}$ reflecting roughly a factor of two larger cross section than $Z^{\prime}$ caused primarily by the relative SM coupling strengths of the two type of bosons to quarks within the SM. The Randall-Sundrum graviton production via gluons pairs is significant at low mass, but becomes negligible at very high mass compared to quark-antiquark production, giving two slopes as a function of integrated luminosity from these two processes and their different parton distribution behavior. Therefore, near the turn-on of the quark-antiquark production process, the increase in the sensitivity for a Randall-Suncrum graviton with integrated luminosity is faster than logarithmic.  None of these characteristics distinguishing the models depend on $\sqrt{s}$, hence Fig.~\ref{figStrongWeak} demonstrates a very similar behavior of the discovery sensitivity among them at four colliders, HL-LHC ($\sqrt{s}=14$ TeV), FNAL-SF or HE-LHC ($\sqrt{s}=27$ TeV), FCC-hh ($\sqrt{s}=100$ TeV) and VLHC ($\sqrt{s} = 300$ TeV).

\begin{figure}[htb]
\begin{center}
\includegraphics[width=0.54\hsize]{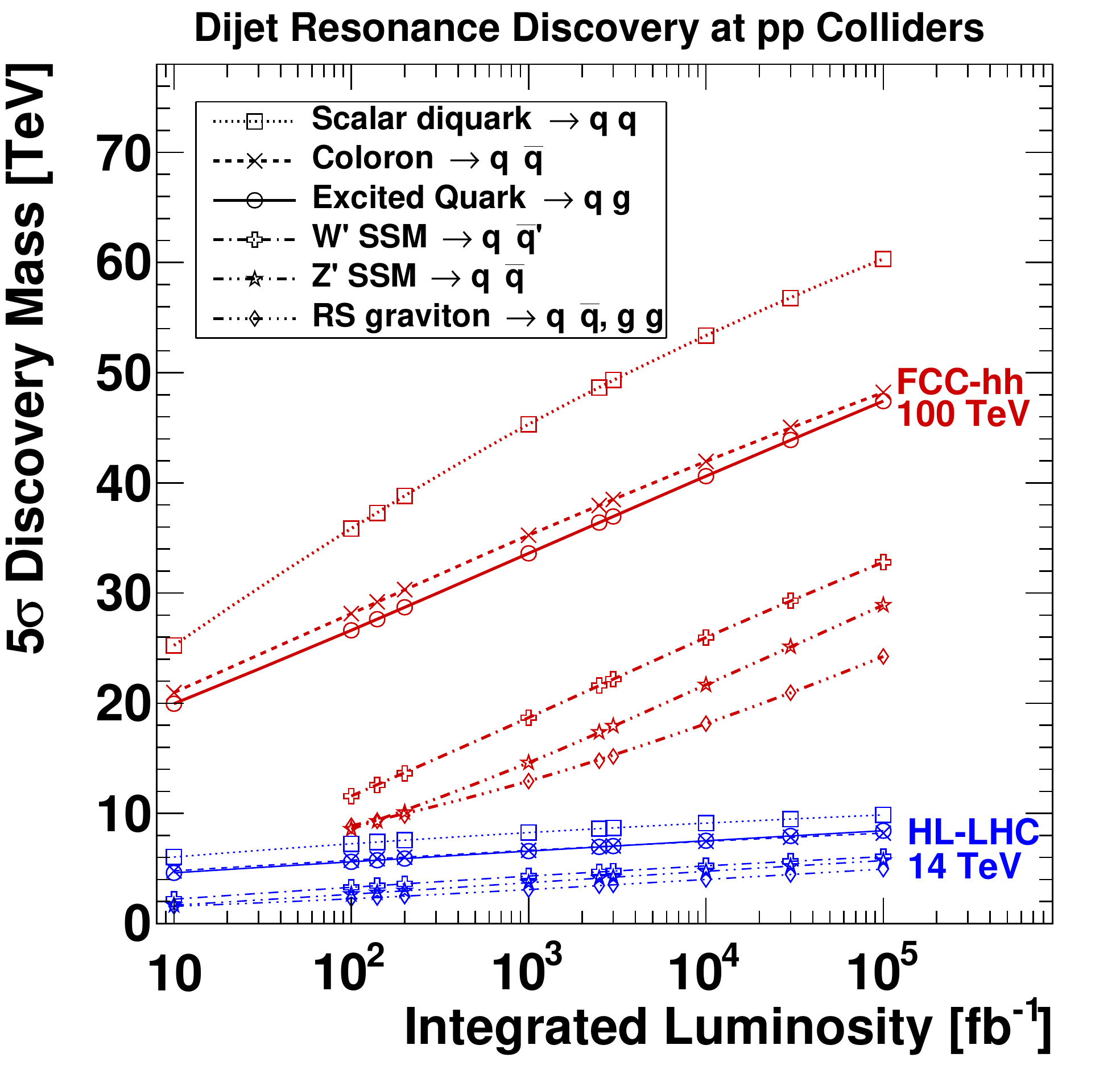}
\includegraphics[width=0.45\hsize]{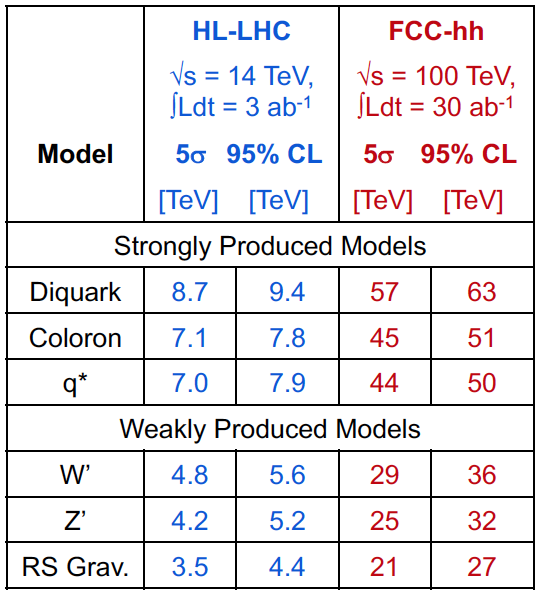}
\end{center}
\caption{Comparison of the sensitivity to dijet resonances of HL-LHC and FCC-hh for six models. Left) The $5\sigma$ discovery sensitivity as a function of integrated luminosity. Right) The $5\sigma$ discovery and 95\% CL exclusion sensitivities for the baseline integrated luminosities of each machine.}
\label{figFCC-LHC}
\end{figure}

\subsection{HL-LHC and FCC-hh}

Two important proton-proton machines to discuss and compare are the HL-LHC and FCC-hh, both based at CERN. The HL-LHC upgrades should be completed over the next several years, and operations are now planned for 2029-42, and hence projections for it are important to understand the sensitivity we can expect to achieve within the next generation. The FCC-hh, while farther off, is being planned and is a critical part of a world-wide strategy for high energy physics~\cite{FCC-hh}. The experimental physics motivation for FCC-hh is clear from Fig.~\ref{figFCC-LHC}. By virtue of the increased collision energy compared to LHC, with sufficient luminosity FCC-hh can proportionally extend the mass reach for discovery of dijet resonances. Fig.~\ref{figFCC-LHC} demonstrates that the mass reach of FCC-hh with 30 ab$^{-1}$ is roughly six times that of HL-LHC with 3 ab$^{-1}$.  The advantage of energy is realized soon after turning on the machine.  With just 10 fb$^{-1}$ of integrated luminosity, accumulated during the first year of data taking, the mass reach for strongly
produced resonances like a coloron or an excited quark, is already three times that of HL-LHC with a full 3 ab$^{-1}$. A modest additional mass reach beyond the baseline can be achieved by running at higher luminosity. For this phenomena, with 100 ab$^{-1}$ at FCC-hh, a factor of seven increase in mass reach is achieved over HL-LHC, fully proportional to the increase in collision energy. Fig.~\ref{figFCC-LHC} also makes clear the significant difference in sensitivity between the three models we classify as strongly produced, and the three that are weakly produced. We see that for FCC-hh we would be able to discover strongly produced resonances up to about one half the $\sqrt{s}$, and weakly produced resonances up to about a quarter of the $\sqrt{s}$, and a little beyond those fractions of $\sqrt{s}$ for HL-LHC because it will accumulate a high luminosity sample.

\subsection{Sensitivities of Colliders}

\begin{figure}[htb]
\begin{center}
\includegraphics[width=\hsize]{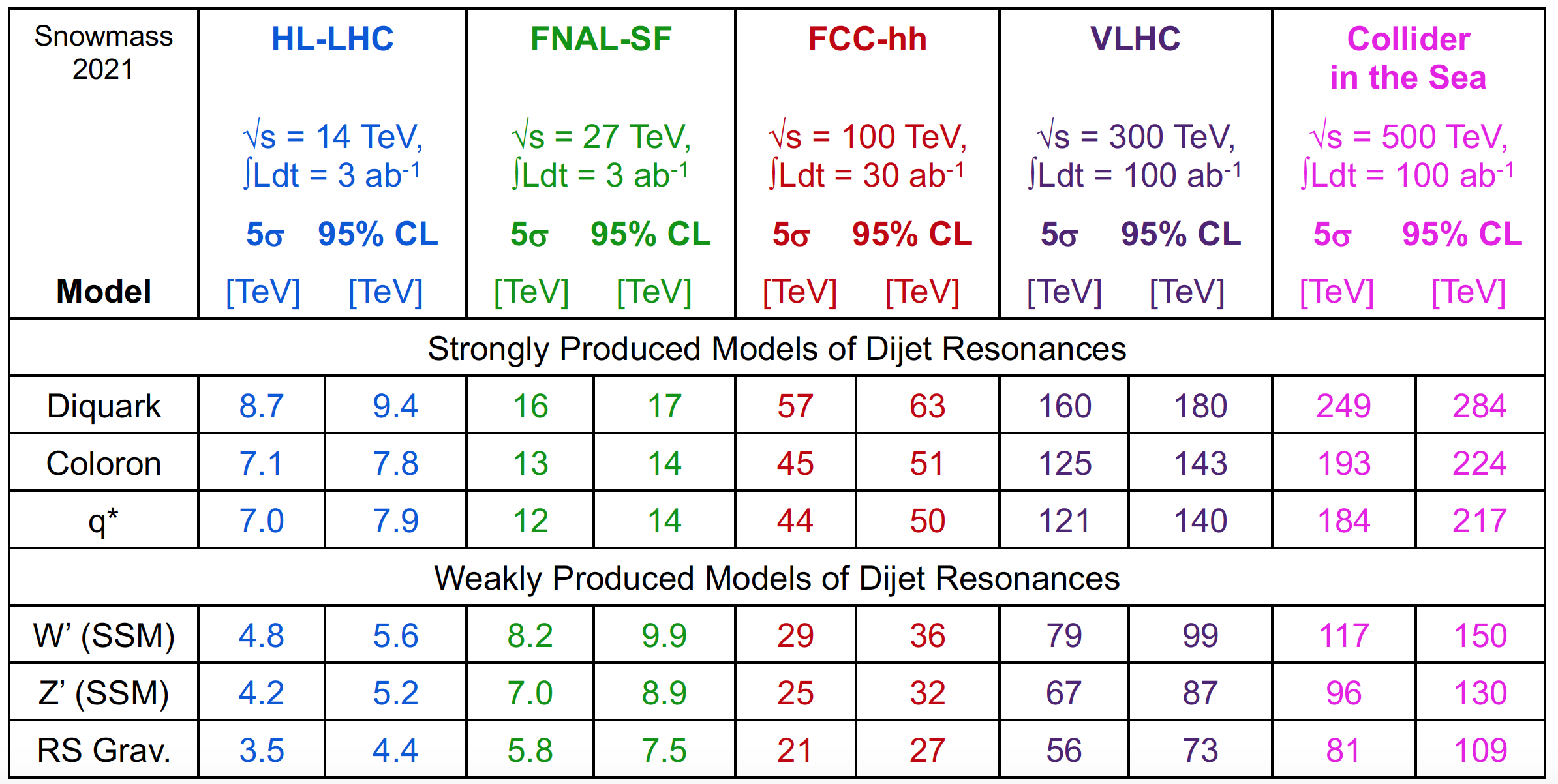}
\end{center}
\caption{Comparison of the sensitivity to dijet resonances of five pp Colliders at their approximate baseline integrated luminosities. The mass for discovery at $5\sigma$, or exclusion at 95\% confidence level, is listed  for six models in descending order of model cross section. }
\label{figSensitivityTable}
\end{figure}

In Fig.~\ref{figSensitivityTable} we summarize the sensitivity of the five major options for future colliders: HL-LHC, FNAL-SF, FCC-hh at default energy $\sqrt{s}=100$ TeV, VLHC and the Collider in the Sea. Baseline integrated luminosities for HL-LHC and FCC-hh are pretty well established, but the baseline integrated luminosities listed for the other collider options are only a rough guess, and are limited by the integrated luminosity choices we made when we began the study.  In the appendix in Tables~\ref{tabDiscoveryStrong}-\ref{tabLimitWeak} we present the sensitivities for all eight collision energies and ten integrated luminosities, along with a few remarks on the range of applicability. The production strength is useful for organizing the results, and we again note that very roughly speaking, the discovery mass reach of a proton-proton collider is about half it's collision energy for strongly produced dijet resonances and about one quarter of it's collision energy for weakly produced dijet resonances.

\section{Conclusions}
\label{secConclusions}

Dijet resonances are a powerful channel for discovery at pp colliders, that directly explores the highest mass scales of new physics in parton-parton collisions. Dijet resonance searches are sensitive to heavy bosons and fermions, strongly or weakly produced. We have estimated the sensitivity for discovery or exclusion to multiple models of dijet resonances, across a wide range of pp collision energy ($\sqrt{s}$) and integrated luminosity ($\int L\/dt$). Sensitivity to dijet resonance production at pp colliders scales as expected, increasing roughly linearly with $\sqrt{s}$ and logarithmically with $\int L\/dt$.  A very rough rule of thumb is that the sensitivity is approximately half the collision energy for strongly produced models and one quarter the collision energy for weakly produced models. These results may be compared with those from other channels, and at other types of future colliders.

\section{Appendix: Sensitivity Tables}
\label{secAppendix}
The full results of the study are shown in Tables~\ref{tabDiscoveryStrong}-\ref{tabLimitWeak}. We have produced results for the full matrix of eight collider energies and ten integrated luminosities.

There are some exceptions and caveats we want to note. Combinations where there is no entry in the table, are those for which the sensitivities (or signal windows) are outside of the range of diparton mass for which we calculated the QCD background: $0.05< m/\sqrt{s} < 0.75$. Trivially, some of the integrated luminosities are unlikely to be achieved at a given machine, and are only listed for completeness, like $10^5$ fb$^{-1}$ for $\sqrt{s}=13$ TeV. More important, there is a maximum resonance mass for which our approximation of constant window acceptance is reasonable.  For example, for excited quarks, the acceptance for our window in Fig.~\ref{figAcceptance} is fairly flat for excited quarks with mass $M/\sqrt{s} < 0.5$. However, at higher values of resonance mass a very long tail to lower dijet masses is significantly reducing the acceptance, cutting the acceptance in half when the resonance mass is $M/\sqrt{s}=0.6$. We estimate that this reduction in acceptance leads to roughly a 6\% reduction in sensitivity for excited quarks at $M/\sqrt{s}=0.6$.  So, for example, the HL-LHC discovery mass of 7 TeV for an excited quark at the baseline integrated luminosity of 3 ab$^{-1}$ in Table~\ref{tabDiscoveryStrong} is just at the edge of the region of flat acceptance, $M/\sqrt{s} < 0.5$, and is likely accurate. However, the HL-LHC mass limit of 7.9 TeV at 3 ab$^{-1}$ in Table~\ref{tabLimitStrong} corresponds to $M/\sqrt{s} = 0.56 > 0.5$, which is outside the range of flat acceptance. We estimate this 7.9 TeV sensitivity should be reduced by roughly 0.3 TeV to account for the estimated reduction in acceptance. While this is still reasonable accuracy, at higher values of integrated luminosity, well beyond the HL-LHC baseline, the quoted sensitivity needs to be reduced significantly more. Other machines run at ultra-high integrated luminosity, much more than their baseline, may also be sensitive to excited quark masses where our acceptance approximation is no longer valid and the mass sensitivities quoted would not be accurate. Similar statements may apply to the other strongly produced models, although the models that couple to gluons have more serious problems with acceptance at ultra-high resonance masses. 

\begin{table}[hbt]
\centering
\begin{small}
\begin{tabular}{|r|rrrrrrrr|} \hline
 \multicolumn{9}{|c|}{ \textbf{5$\sigma$ discovery mass of strong dijet resonances (TeV)}} \\ \hline
 $\int L\/dt$ & \multicolumn{8}{c|}{pp Collider $\sqrt{s}$ (TeV)}  \\ 
 (fb$^{-1})$  & 13 & 14 & 27 & 75 & 100 & 150 & 300 & 500  \\ \cline{2-9}
  & \multicolumn{8}{c|}{\textbf{Scalar diquark $\rightarrow qq$} } \\ \hline
 10$^{1}$             &   5.7 &   6.0 &  10.1 &  20.9 &  25.2 &  31.9 &  44.3 &  52.0 \\
 10$^{2}$             &   6.8 &   7.2 &  12.6 &  28.8 &  35.9 &  48.2 &  77.1 & 104.5 \\
 $1.4 \times 10^{2}$  &   6.9 &   7.4 &  13.0 &  29.9 &  37.3 &  50.5 &  81.7 & 112.1 \\
 $2.0 \times 10^{2}$  &   7.1 &   7.6 &  13.3 &  31.0 &  38.8 &  52.9 &  86.7 & 120.8 \\
 10$^{3}$             &   7.7 &   8.3 &  14.8 &  35.6 &  45.3 &  62.9 & 108.1 & 157.3 \\
 $2.5 \times 10^{3}$  &   8.0 &   8.6 &  15.5 &  38.1 &  48.7 &  68.4 & 119.9 & 177.5 \\
 $3.0 \times 10^{3}$  &   8.1 &   8.7 &  15.7 &  38.5 &  49.3 &  69.3 & 121.9 & 181.2 \\
 10$^{4}$             &   8.5 &   9.1 &  16.6 &  41.5 &  53.4 &  75.9 & 136.1 & 205.9 \\
 $3.0 \times 10^{4}$  &   8.8 &   9.5 &  17.4 &  43.9 &  56.8 &  81.3 & 148.0 & 227.4 \\
 10$^{5}$             &   9.2 &   9.9 &  18.2 &  46.4 &  60.3 &  86.7 & 160.0 & 249.2 \\
\hline
  & \multicolumn{8}{c|}{\textbf{Coloron $\rightarrow q\bar{q}$} } \\ \hline
 10$^{1}$             &   4.5 &   4.8 &   8.0 &  17.1 &  21.0 &  27.6 &  42.8 &  57.5 \\
 10$^{2}$             &   5.4 &   5.7 &   9.9 &  22.6 &  28.1 &  38.2 &  62.8 &  88.7 \\
 $1.4 \times 10^{2}$  &   5.5 &   5.9 &  10.2 &  23.3 &  29.2 &  39.7 &  65.8 &  93.7 \\
 $2.0 \times 10^{2}$  &   5.7 &   6.0 &  10.5 &  24.1 &  30.3 &  41.4 &  69.1 &  99.1 \\
 10$^{3}$             &   6.2 &   6.6 &  11.7 &  27.8 &  35.2 &  48.8 &  84.0 & 123.5 \\
 $2.5 \times 10^{3}$  &   6.5 &   7.0 &  12.4 &  29.9 &  37.9 &  53.1 &  92.5 & 137.4 \\
 $3.0 \times 10^{3}$  &   6.6 &   7.1 &  12.6 &  30.2 &  38.5 &  53.9 &  94.2 & 140.2 \\
 10$^{4}$             &   7.0 &   7.5 &  13.4 &  32.8 &  42.0 &  59.2 & 105.2 & 158.6 \\
 $3.0 \times 10^{4}$  &   7.3 &   7.8 &  14.2 &  35.0 &  45.0 &  63.8 & 114.8 & 175.4 \\
 10$^{5}$             &   7.7 &   8.2 &  14.9 &  37.4 &  48.2 &  68.8 & 125.2 & 192.9 \\
\hline
 & \multicolumn{8}{c|}{\textbf{Excited quark $\rightarrow qg$} } \\ \hline
 10$^{1}$             &   4.3 &   4.6 &   7.7 &  16.3 &  20.0 &  26.4 &  41.7 &  57.3 \\
 10$^{2}$             &   5.3 &   5.6 &   9.6 &  21.4 &  26.6 &  36.0 &  59.2 &  84.1 \\
 $1.4 \times 10^{2}$  &   5.4 &   5.7 &   9.9 &  22.2 &  27.6 &  37.5 &  62.0 &  88.4 \\
 $2.0 \times 10^{2}$  &   5.5 &   5.9 &  10.2 &  23.0 &  28.7 &  39.0 &  65.0 &  93.1 \\
 10$^{3}$             &   6.2 &   6.6 &  11.5 &  26.7 &  33.6 &  46.3 &  78.9 & 115.4 \\
 $2.5 \times 10^{3}$  &   6.5 &   7.0 &  12.2 &  28.8 &  36.4 &  50.5 &  87.2 & 128.6 \\
 $3.0 \times 10^{3}$  &   6.6 &   7.0 &  12.4 &  29.2 &  37.0 &  51.3 &  88.8 & 131.3 \\
 10$^{4}$             &   7.0 &   7.5 &  13.3 &  31.9 &  40.6 &  56.8 &  99.7 & 149.3 \\
 $3.0 \times 10^{4}$  &   7.4 &   8.0 &  14.2 &  34.4 &  43.9 &  61.8 & 109.7 & 165.8 \\
 10$^{5}$             &   7.9 &   8.4 &  15.1 &  37.0 &  47.4 &  67.2 & 120.6 & 184.0 \\
\hline
\end{tabular}
\end{small}
\caption{ Discovery sensitivity for strongly produced dijet resonances at proton-proton colliders. As a function of collision energy (columns) and integrated luminosity (rows) the largest resonance mass for which a $5\sigma$ discovery can be expected is listed for three strongly produced models: diquarks (upper), colorons (middle) and excited quarks (lower)}
\label{tabDiscoveryStrong}
\end{table}

\clearpage

\begin{table}[hbt]
\centering
\begin{small}
\begin{tabular}{|r|rrrrrrrr|} \hline
 \multicolumn{9}{|c|}{ \textbf{5$\sigma$ discovery mass of weak dijet resonances (TeV)}} \\ \hline
 $\int L\/dt$ & \multicolumn{8}{c|}{pp Collider $\sqrt{s}$ (TeV)}  \\ 
 (fb$^{-1})$  & 13 & 14 & 27 & 75 & 100 & 150 & 300 & 500  \\ \cline{2-9}
  & \multicolumn{8}{c|}{\textbf{W$^{\prime}$ SSM $\rightarrow q\bar{q}^{\prime}$} } \\ \hline
 10$^{1}$             &   2.1 &   2.2 &   3.2 &   5.3 &  --- &   --- &   --- &   --- \\
 10$^{2}$             &   3.1 &   3.3 &   5.2 &   9.9 &  11.6 &  14.2 &  19.7 &  --- \\
 $1.4 \times 10^{2}$  &   3.3 &   3.4 &   5.5 &  10.6 &  12.6 &  15.5 &  22.0 &  --- \\
 $2.0 \times 10^{2}$  &   3.4 &   3.6 &   5.8 &  11.5 &  13.6 &  17.1 &  24.4 &  30.1 \\
 10$^{3}$             &   4.1 &   4.3 &   7.2 &  15.3 &  18.7 &  24.3 &  36.8 &  48.0 \\
 $2.5 \times 10^{3}$  &   4.4 &   4.7 &   8.0 &  17.5 &  21.6 &  28.7 &  44.8 &  60.6 \\
 $3.0 \times 10^{3}$  &   4.5 &   4.8 &   8.2 &  18.0 &  22.2 &  29.6 &  46.5 &  63.2 \\
 10$^{4}$             &   4.9 &   5.2 &   9.1 &  20.8 &  26.0 &  35.2 &  57.8 &  81.0 \\
 $3.0 \times 10^{4}$  &   5.3 &   5.7 &   9.9 &  23.2 &  29.3 &  40.3 &  68.1 &  98.1 \\
 10$^{5}$             &   5.7 &   6.1 &  10.8 &  25.8 &  32.8 &  45.7 &  79.2 & 116.6 \\
\hline
  & \multicolumn{8}{c|}{\textbf{Z$^{\prime}$ SSM $\rightarrow q\bar{q}$}  } \\ \hline
 10$^{1}$             &   1.6 &   1.7 &   2.4 &   --- &   --- &   --- &   --- &   --- \\
 10$^{2}$             &   2.5 &   2.7 &   4.1 &   7.3 &   8.6 &  10.5 &   --- &   --- \\
 $1.4 \times 10^{2}$  &   2.7 &   2.8 &   4.4 &   7.9 &   9.3 &  11.5 &   --- &   --- \\
 $2.0 \times 10^{2}$  &   2.8 &   3.0 &   4.7 &   8.7 &  10.1 &  12.7 &   --- &   --- \\
 10$^{3}$             &   3.5 &   3.7 &   6.0 &  12.2 &  14.6 &  18.7 &  27.4 &  35.8 \\
 $2.5 \times 10^{3}$  &   3.9 &   4.1 &   6.8 &  14.3 &  17.4 &  22.5 &  34.2 &  45.3 \\
 $3.0 \times 10^{3}$  &   4.0 &   4.2 &   7.0 &  14.7 &  17.9 &  23.4 &  35.7 &  47.2 \\
 10$^{4}$             &   4.5 &   4.7 &   8.0 &  17.6 &  21.7 &  28.9 &  45.6 &  62.7 \\
 $3.0 \times 10^{4}$  &   4.9 &   5.2 &   9.0 &  20.2 &  25.1 &  34.0 &  55.6 &  77.9 \\
 10$^{5}$             &   5.4 &   5.7 &  10.0 &  23.0 &  28.9 &  39.6 &  66.6 &  95.9 \\
\hline
 & \multicolumn{8}{c|}{\textbf{RS graviton $\rightarrow q\bar{q}$,gg  } } \\ \hline
 10$^{1}$             &   1.5 &   1.6 &   2.5 &   4.9 &   --- &   --- &   --- &   --- \\
 10$^{2}$             &   2.1 &   2.2 &   3.6 &   7.2 &   8.9 &  11.8 &  18.3 &   --- \\
 $1.4 \times 10^{2}$  &   2.2 &   2.4 &   3.8 &   7.6 &   9.3 &  12.4 &  19.6 &   --- \\
 $2.0 \times 10^{2}$  &   2.4 &   2.5 &   4.0 &   8.1 &   9.8 &  13.2 &  21.0 &   --- \\
 10$^{3}$             &   2.9 &   3.1 &   5.0 &  10.5 &  12.9 &  17.1 &  27.4 &  39.2 \\
 $2.5 \times 10^{3}$  &   3.2 &   3.4 &   5.7 &  12.1 &  14.8 &  19.8 &  31.6 &  45.3 \\
 $3.0 \times 10^{3}$  &   3.3 &   3.5 &   5.8 &  12.4 &  15.2 &  20.4 &  32.7 &  46.5 \\
 10$^{4}$             &   3.8 &   4.0 &   6.7 &  14.6 &  18.1 &  24.3 &  39.9 &  56.7 \\
 $3.0 \times 10^{4}$  &   4.2 &   4.5 &   7.6 &  16.9 &  21.0 &  28.4 &  46.9 &  67.7 \\
 10$^{5}$             &   4.6 &   4.9 &   8.5 &  19.4 &  24.2 &  33.1 &  55.7 &  80.7 \\
\hline
\end{tabular}
\end{small}
\caption{ Discovery sensitivity for weakly produced dijet resonances at proton-proton colliders. As a function of collision energy (columns) and integrated luminosity (rows) the largest resonance mass for which a $5\sigma$ discovery can be expected is listed for three strongly produced models: heavy bosons $W^{\prime}$ (upper), $Z^{\prime}$ (middle) and Randall-Sundrum graviton (lower)}
\label{tabDiscoveryWeak}
\end{table}

\clearpage

\begin{table}[hbt]
\centering
\begin{small}
\begin{tabular}{|r|rrrrrrrr|} \hline
 \multicolumn{9}{|c|}{ \textbf{95\% CL mass limit of strong dijet resonances (TeV)}} \\ \hline
 $\int L\/dt$ & \multicolumn{8}{c|}{pp Collider $\sqrt{s}$ (TeV)}  \\ 
 (fb$^{-1})$  & 13 & 14 & 27 & 75 & 100 & 150 & 300 & 500  \\ \cline{2-9}
  & \multicolumn{8}{c|}{\textbf{Scalar diquark $\rightarrow qq$} } \\ \hline
 10$^{1}$             &   6.7 &   7.2 &  12.5 &  28.4 &  35.3 &  47.3 &  75.5 & 101.8 \\
 10$^{2}$             &   7.7 &   8.2 &  14.7 &  35.3 &  44.9 &  62.2 & 106.5 & 154.4 \\
 $1.4 \times 10^{2}$  &   7.8 &   8.4 &  15.0 &  36.2 &  46.1 &  64.2 & 110.6 & 161.6 \\
 $2.0 \times 10^{2}$  &   7.9 &   8.5 &  15.3 &  37.2 &  47.4 &  66.3 & 115.2 & 169.6 \\
 10$^{3}$             &   8.5 &   9.1 &  16.5 &  41.3 &  53.0 &  75.3 & 134.7 & 203.4 \\
 $2.5 \times 10^{3}$  &   8.8 &   9.4 &  17.2 &  43.3 &  55.9 &  79.8 & 144.6 & 221.4 \\
 $3.0 \times 10^{3}$  &   8.8 &   9.4 &  17.3 &  43.7 &  56.5 &  80.7 & 146.6 & 225.1 \\
 10$^{4}$             &   9.2 &   9.8 &  18.1 &  46.2 &  60.0 &  86.2 & 158.8 & 246.8 \\
 $3.0 \times 10^{4}$  &   9.5 &  10.2 &  18.8 &  48.4 &  62.9 &  91.0 & 169.2 & 265.0 \\
 10$^{5}$             &   --- &   --- &  19.5 &  50.5 &  65.9 &  95.7 & 180.0 & 283.8 \\
\hline
  & \multicolumn{8}{c|}{\textbf{Coloron $\rightarrow q\bar{q}$} } \\ \hline
 10$^{1}$             &   5.4 &   5.7 &   9.9 &  22.3 &  27.8 &  37.7 &  61.9 &  87.1 \\
 10$^{2}$             &   6.2 &   6.6 &  11.7 &  27.6 &  35.0 &  48.4 &  83.1 & 121.8 \\
 $1.4 \times 10^{2}$  &   6.3 &   6.7 &  11.9 &  28.4 &  36.0 &  50.0 &  86.2 & 127.0 \\
 $2.0 \times 10^{2}$  &   6.4 &   6.9 &  12.2 &  29.2 &  37.0 &  51.6 &  89.6 & 132.4 \\
 10$^{3}$             &   7.0 &   7.4 &  13.4 &  32.6 &  41.7 &  58.8 & 104.3 & 157.1 \\
 $2.5 \times 10^{3}$  &   7.2 &   7.7 &  14.0 &  34.5 &  44.3 &  62.7 & 112.4 & 171.1 \\
 $3.0 \times 10^{3}$  &   7.3 &   7.8 &  14.1 &  34.9 &  44.8 &  63.5 & 114.0 & 173.9 \\
 10$^{4}$             &   7.6 &   8.2 &  14.9 &  37.2 &  48.0 &  68.5 & 124.4 & 191.5 \\
 $3.0 \times 10^{4}$  &   8.0 &   8.5 &  15.6 &  39.3 &  50.8 &  72.8 & 133.7 & 207.3 \\
 10$^{5}$             &   8.3 &   8.9 &  16.3 &  41.4 &  53.7 &  77.4 & 143.2 & 224.1 \\
\hline
 & \multicolumn{8}{c|}{\textbf{Excited quark $\rightarrow qg$} } \\ \hline
 10$^{1}$             &   5.2 &   5.6 &   9.5 &  21.2 &  26.3 &  35.6 &  58.4 &  82.7 \\
 10$^{2}$             &   6.1 &   6.5 &  11.4 &  26.5 &  33.3 &  45.9 &  78.1 & 113.9 \\
 $1.4 \times 10^{2}$  &   6.3 &   6.7 &  11.7 &  27.2 &  34.4 &  47.4 &  81.1 & 118.8 \\
 $2.0 \times 10^{2}$  &   6.4 &   6.8 &  12.0 &  28.0 &  35.5 &  49.1 &  84.3 & 124.0 \\
 10$^{3}$             &   7.0 &   7.5 &  13.3 &  31.7 &  40.4 &  56.4 &  98.9 & 147.9 \\
 $2.5 \times 10^{3}$  &   7.4 &   7.9 &  14.0 &  33.8 &  43.1 &  60.6 & 107.2 & 161.6 \\
 $3.0 \times 10^{3}$  &   7.4 &   7.9 &  14.1 &  34.2 &  43.6 &  61.4 & 108.9 & 164.4 \\
 10$^{4}$             &   7.9 &   8.4 &  15.0 &  36.8 &  47.2 &  66.8 & 119.8 & 182.6 \\
 $3.0 \times 10^{4}$  &   8.3 &   8.8 &  15.9 &  39.2 &  50.4 &  71.6 & 129.6 & 199.2 \\
 10$^{5}$             &   8.7 &   9.3 &  16.8 &  41.7 &  53.8 &  76.8 & 140.2 & 217.0 \\
\hline
\end{tabular}
\end{small}
\caption{ Mass limit sensitivity for strongly produced dijet resonances at proton-proton colliders. As a function of collision energy (columns) and integrated luminosity (rows) the largest resonance mass for which a 95\% CL exclusion can be expected is listed for three strongly produced models: diquarks (upper), colorons (middle) and excited quarks (lower)}
\label{tabLimitStrong}
\end{table}

\clearpage

\begin{table}[hbt]
\centering
\begin{small}
\begin{tabular}{|r|rrrrrrrr|} \hline
 \multicolumn{9}{|c|}{ \textbf{95\% CL mass limit of weak dijet resonances (TeV)} } \\ \hline
 $\int L\/dt$ & \multicolumn{8}{c|}{pp Collider $\sqrt{s}$ (TeV)}  \\ 
 (fb$^{-1})$  & 13 & 14 & 27 & 75 & 100 & 150 & 300 & 500  \\ \cline{2-9}
  & \multicolumn{8}{c|}{\textbf{W$^{\prime}$ SSM $\rightarrow q\bar{q}^{\prime}$} } \\ \hline
 10$^{1}$             &   3.1 &   3.3 &   5.2 &   9.7 &  11.3 &  14.0 &  19.2 &   --- \\
 10$^{2}$             &   4.0 &   4.3 &   7.2 &  15.2 &  18.4 &  24.0 &  36.2 &  47.2 \\
 $1.4 \times 10^{2}$  &   4.2 &   4.4 &   7.5 &  16.0 &  19.5 &  25.6 &  39.1 &  51.2 \\
 $2.0 \times 10^{2}$  &   4.3 &   4.6 &   7.8 &  16.8 &  20.7 &  27.3 &  42.2 &  56.3 \\
 10$^{3}$             &   4.9 &   5.2 &   9.0 &  20.6 &  25.7 &  34.9 &  57.1 &  79.8 \\
 $2.5 \times 10^{3}$  &   5.2 &   5.6 &   9.7 &  22.7 &  28.5 &  39.1 &  65.7 &  94.0 \\
 $3.0 \times 10^{3}$  &   5.3 &   5.6 &   9.9 &  23.1 &  29.1 &  40.0 &  67.4 &  96.9 \\
 10$^{4}$             &   5.7 &   6.0 &  10.8 &  25.7 &  32.6 &  45.4 &  78.5 & 115.4 \\
 $3.0 \times 10^{4}$  &   6.0 &   6.4 &  11.5 &  27.9 &  35.7 &  50.1 &  88.3 & 132.1 \\
 10$^{5}$             &   6.4 &   6.8 &  12.3 &  30.3 &  38.9 &  55.1 &  98.6 & 150.0 \\
 \hline
  & \multicolumn{8}{c|}{\textbf{Z$^{\prime}$ SSM $\rightarrow q\bar{q}$}  } \\ \hline
 10$^{1}$             &   2.5 &   2.6 &   4.0 &   7.2 &   8.4 &  10.2 &   --- &   --- \\
 10$^{2}$             &   3.5 &   3.7 &   6.0 &  12.0 &  14.4 &  18.4 &  27.0 &  35.0 \\
 $1.4 \times 10^{2}$  &   3.6 &   3.8 &   6.3 &  12.8 &  15.4 &  19.8 &  29.1 &  38.5 \\
 $2.0 \times 10^{2}$  &   3.8 &   4.0 &   6.6 &  13.6 &  16.5 &  21.3 &  31.8 &  42.2 \\
 10$^{3}$             &   4.4 &   4.7 &   8.0 &  17.4 &  21.4 &  28.5 &  45.0 &  61.7 \\
 $2.5 \times 10^{3}$  &   4.8 &   5.1 &   8.8 &  19.6 &  24.3 &  32.8 &  53.2 &  74.2 \\
 $3.0 \times 10^{3}$  &   4.9 &   5.2 &   8.9 &  20.0 &  24.9 &  33.6 &  54.9 &  76.8 \\
 10$^{4}$             &   5.3 &   5.7 &   9.9 &  22.8 &  28.7 &  39.3 &  65.9 &  94.7 \\
 $3.0 \times 10^{4}$  &   5.7 &   6.1 &  10.8 &  25.4 &  32.1 &  44.4 &  76.1 & 111.4 \\
 10$^{5}$             &   6.2 &   6.6 &  11.7 &  28.1 &  35.7 &  49.9 &  87.3 & 129.9 \\
\hline
 & \multicolumn{8}{c|}{\textbf{RS graviton $\rightarrow q\bar{q}$,gg  } } \\ \hline
 10$^{1}$             &   2.1 &   2.2 &   3.6 &   7.1 &   8.8 &  11.6 &  18.0 &   --- \\
 10$^{2}$             &   2.9 &   3.1 &   5.0 &  10.4 &  12.8 &  16.8 &  27.1 &  38.8 \\
 $1.4 \times 10^{2}$  &   3.0 &   3.2 &   5.2 &  10.9 &  13.5 &  17.8 &  28.5 &  41.0 \\
 $2.0 \times 10^{2}$  &   3.1 &   3.3 &   5.5 &  11.5 &  14.2 &  18.9 &  29.9 &  43.3 \\
 10$^{3}$             &   3.7 &   4.0 &   6.7 &  14.5 &  18.0 &  24.0 &  39.5 &  56.0 \\
 $2.5 \times 10^{3}$  &   4.1 &   4.3 &   7.4 &  16.3 &  20.3 &  27.4 &  45.0 &  65.1 \\
 $3.0 \times 10^{3}$  &   4.2 &   4.4 &   7.5 &  16.7 &  20.7 &  28.1 &  46.3 &  66.9 \\
 10$^{4}$             &   4.6 &   4.9 &   8.5 &  19.2 &  24.0 &  32.8 &  55.2 &  79.8 \\
 $3.0 \times 10^{4}$  &   5.0 &   5.4 &   9.3 &  21.6 &  27.1 &  37.3 &  63.6 &  93.3 \\
 10$^{5}$             &   5.5 &   5.8 &  10.3 &  24.1 &  30.6 &  42.4 &  73.4 & 108.8 \\ 
\hline
\end{tabular}
\end{small}
\caption{ Mass limit sensitivity for weakly produced dijet resonances at proton-proton colliders. As a function of collision energy (columns) and integrated luminosity (rows) the largest resonance mass for which a 95\% CL exclusion  can be expected is listed for three strongly produced models: heavy bosons $W^{\prime}$ (upper), $Z^{\prime}$ (middle) and Randall-Sundrum graviton (lower)}
\label{tabLimitWeak}
\end{table}

\clearpage

\noindent\textbf{\Large Acknowledgements}

\noindent Work supported by the Fermi National Accelerator Laboratory, managed and operated by Fermi Research Alliance, LLC under Contract No. DE-AC02-07CH11359 with the U.S. Department of Energy. The U.S. Government retains and the publisher, by accepting the article for publication, acknowledges that the U.S. Government retains a non-exclusive, paid-up, irrevocable, world-wide license to publish or reproduce the published form of this manuscript, or allow others to do so, for U.S. Government purposes.  This work was also supported by the Turkish Energy, Nuclear and Mineral Research Agency.







\end{document}